\documentclass{aastex631}

\usepackage{amsmath}
\usepackage{graphicx}
\usepackage{subfigure}
\usepackage{fourier}
\usepackage{arcs}
\usepackage{url}
\usepackage{color}
\usepackage{amsbsy}
\usepackage{mathrsfs}
\usepackage{bm}
\usepackage{booktabs}
\usepackage{multirow}
\usepackage{url}
\usepackage{ulem}
\usepackage{tikz}
\begin{document}

\title{SIP-IFVM: A time-evolving coronal model with an extended magnetic field decomposition strategy}

\author[0000-0002-4217-6990]{Hao P. Wang}
\affiliation{Centre for Mathematical Plasma-Astrophysics, Department of Mathematics, KU Leuven, Celestijnenlaan 200B, 3001 Leuven, Belgium}

\author[0000-0003-4716-2958]{Li P. Yang}
\affiliation{State Key Laboratory of Space Weather, Chinese Academy of Sciences, Beijing 100190, China; \url{lpyang@spaceweather.ac.cn};
\url{haopeng.wang1@kuleuven.be}}

\author[0000-0002-1743-0651]{S. Poedts}
\affiliation{Centre for Mathematical Plasma-Astrophysics, Department of Mathematics, KU Leuven, Celestijnenlaan 200B, 3001 Leuven, Belgium}
\affiliation{Institute of Physics, University of Maria Curie-Skłodowska, ul. Radziszewskiego 10, 20-031 Lublin, Poland}

\author[0000-0003-4017-215X]{A. Lani}
\affiliation{Centre for Mathematical Plasma-Astrophysics, Department of Mathematics, KU Leuven, Celestijnenlaan 200B, 3001 Leuven, Belgium}
\affiliation{Von Karman Institute For Fluid Dynamics, Waterloosesteenweg 72, 1640 Sint-Genesius-Rode, Brussels, Belgium}

\author[0000-0002-4391-393X]{Yu H. Zhou}
\affiliation{Centre for Mathematical Plasma-Astrophysics, Department of Mathematics, KU Leuven, Celestijnenlaan 200B, 3001 Leuven, Belgium}

\author{Yu H. Gao}
\affiliation{Centre for Mathematical Plasma-Astrophysics, Department of Mathematics, KU Leuven, Celestijnenlaan 200B, 3001 Leuven, Belgium}
\affiliation{School of Earth and Space Sciences, Peking University, Beijing 100871, China}

\author[0000-0002-4014-1815]{L. Linan}
\affiliation{Centre for Mathematical Plasma-Astrophysics, Department of Mathematics, KU Leuven, Celestijnenlaan 200B, 3001 Leuven, Belgium}

\author{Jia K. Lv}
\affiliation{Beijing Institute of Applied Meteorology, Beijing 100029, China}

\author[0000-0002-1986-4496]{T. Baratashvili}
\affiliation{Centre for Mathematical Plasma-Astrophysics, Department of Mathematics, KU Leuven, Celestijnenlaan 200B, 3001 Leuven, Belgium}

\author[0000-0002-4205-5566]{Jing H. Guo}
\affiliation{Centre for Mathematical Plasma-Astrophysics, Department of Mathematics, KU Leuven, Celestijnenlaan 200B, 3001 Leuven, Belgium}
\affiliation{School of Astronomy and Space Science and Key Laboratory of Modern Astronomy and Astrophysics, Nanjing University, Nanjing 210023, China}

\author[0000-0001-7655-5000]{L. Rong}
\affiliation{Centre for Mathematical Plasma-Astrophysics, Department of Mathematics, KU Leuven, Celestijnenlaan 200B, 3001 Leuven, Belgium}
\affiliation{School of Earth and Space Sciences, Peking University, Beijing 100871, China}

\author{Z. Su}
\affiliation{SIGMA Weather Group, State Key Laboratory for Space Weather, National Space Science Center, Chinese Academy of Sciences, Beijing 100190, People's
Republic of China}

\author{Cai X. Li}
\affiliation{Shenzhen Key Laboratory of Ultraintense Laser and Advanced Material Technology, Center for Intense Laser Application Technology, and College of Engineering Physics, Shenzhen Technology University, Shenzhen 518118, China}

\author{M. Zhang}
\affiliation{State Key Laboratory of Space Weather, Chinese Academy of Sciences, Beijing 100190, China}

\author[0000-0001-8495-9179]{Wen W. Wei}
\affiliation{Space Sciences Laboratory, University of California, Berkeley, CA 94720, USA}

\author{Y. Yang}
\affiliation{School of Mathematical Sciences, Nanjing Normal University, Nanjing 210023, People's Republic of China}
\affiliation{Ministry of Education Key Laboratory of NSLSCS, People's Republic of China}

\author{Yu C. Li}
\affiliation{Centre for Mathematical Plasma-Astrophysics, Department of Mathematics, KU Leuven, Celestijnenlaan 200B, 3001 Leuven, Belgium}

\author{Xin Y. Ma}
\affiliation{College of Earth and Planetary Sciences, University of Chinese Academy of Sciences, Beijing 100049, People's Republic of China}

\author{E. Husidic}
\affiliation{Centre for Mathematical Plasma-Astrophysics, Department of Mathematics, KU Leuven, Celestijnenlaan 200B, 3001 Leuven, Belgium}

\author[0000-0003-4616-947X]{H.-J. Jeong}
\affiliation{Centre for Mathematical Plasma-Astrophysics, Department of Mathematics, KU Leuven, Celestijnenlaan 200B, 3001 Leuven, Belgium}

\author{Najafi Z. Mahdi}
\affiliation{Centre for Mathematical Plasma-Astrophysics, Department of Mathematics, KU Leuven, Celestijnenlaan 200B, 3001 Leuven, Belgium}

\author{J. Wang}
\affiliation{School of Systems Science, Beijing Normal University, Beijing 100875, China}

\author[0000-0003-3364-9183]{B. Schmieder}
\affiliation{Centre for Mathematical Plasma-Astrophysics, Department of Mathematics, KU Leuven, Celestijnenlaan 200B, 3001 Leuven, Belgium}
\affiliation{Observatoire de Paris, LIRA, UMR8254 (CNRS), F-92195 Meudon Principal Cedex, France}
\affiliation{SUPA, School of Physics $\&$ Astronomy, University of Glasgow, Glasgow G12 8QQ, UK}

\begin{abstract}

  Time-evolving magnetohydrodynamic (MHD) coronal modeling, driven by a series of time-dependent photospheric magnetograms, represents a new generation of coronal simulations. This approach offers greater realism compared to traditional coronal models constrained by a static magnetogram. However, its practical application is seriously limited by low computational efficiency and poor numerical stability.
  Therefore, we propose an extended magnetic field decomposition strategy and implement it in the implicit MHD model to develop a coronal model that is both efficient and numerically stable enough for simulating the long-term evolutions of the global corona.
  The traditional decomposition strategies split the magnetic field into a time-invariant potential field and a time-dependent component $\mathbf{B}_1$. It works well for quasi-steady-state coronal simulations where $\left|\mathbf{B}_1\right|$ is typically small. However, as the inner-boundary magnetic field evolves, $\left|\mathbf{B}_1\right|$ can grow significantly larger and its discretization errors often lead to nonphysical negative thermal pressure, ultimately causing the code to crash. In this paper, we mitigate such undesired situations by introducing a temporally piecewise-constant variable to accommodate part of the non-potential field and remain $\left|\mathbf{B}_1\right|$ consistently small throughout the simulations. We incorporate this novel magnetic field decomposition strategy into our implicit MHD coronal model and apply it to simulate the evolution of coronal structures within 0.1 AU over two solar-maximum Carrington rotations.
  The results show that this coronal model effectively captures observations and performs more than 80 times faster than real time using only 192 CPU cores, making it well-suited for practical applications in simulating the time-evolving corona.

\end{abstract}

%\linenumbers
\section{Introduction}\label{sec:intro}
Space weather refers to the variable physical conditions on the Sun and in the solar wind, magnetosphere, ionosphere, and thermosphere of the Earth, with coronal mass ejections (CMEs), large-scale releases of magnetized plasma structures from the corona into the heliosphere, being one of the primary drivers. Space weather can impact the performance and reliability of both space-borne and ground-based technological systems and pose risks to human health. For instance, the 1989 Quebec blackout was triggered by a geomagnetic storm caused by a CME. The 2003 ``Halloween Storm" damaged satellite electronics due to increased radiation from CMEs. Also, during CME events, the increased radiation exposure raises astronauts' cancer risk. The total economic loss from a single superstorm has been estimated at up to 2.7 $\$$ trillion \citep{Eastwood2017} and the 2019 National Threat and Hazard Identification and Risk Assessment (THIRA) report identified space weather and pandemics as a potential global threat that could disrupt critical infrastructure, including satellite communications, power grids, and other essential systems that society heavily relies on.

To enable timely action in mitigating damage from severe space weather, advanced Sun-to-Earth model chains should be developed \cite[e.g.][]{Feng_2013Chinese,GOODRICH20041469,Hayashi_2021,ODSTRCIL20041311,Pomoell2018020,Poedts_2020,TOTH2012870} to understand space weather mechanisms and provide reliable forecasts hours to days in advance \cite[e.g.][]{BAKER19987,Feng_2011Chinese,Feng_2013Chinese,Feng2020book,Koskinen2017}.
In the Sun-to-Earth modeling chain, observed photospheric magnetic fields serve as input for the solar corona model, which in turn provides inner boundary conditions for the inner heliospheric model. The inner heliospheric model then supplies boundary information to the geomagnetic model. The coronal model is essential for initializing other models and plays a key role in accurately simulating solar disturbances such as CMEs \citep{Brchnelova_2022,Perri_2023,Kuzma_2023}. However, physics-based MHD corona models are also the most complex and computationally intensive component \cite{WangSubmitted,W_SubmittedCOCONUT} and sometimes encounter low-$\beta$ (the ratio of the thermal pressure to the magnetic pressure) problems, where $\beta$ can drop as low as $10^{-4}$ near the solar surface \citep{Bourdin2017}, leading to severe stability and efficiency challenges \citep{Feng_2021,Wang_2022}.

In most MHD coronal simulations, time steps are limited to a few seconds due to the restriction of the Courant-Friedrichs-Lewy (CFL) stability condition. Consequently, depending on the mesh resolution, even state-of-the-art quasi-steady-state coronal simulations require 10$\sim100\;$k CPU-hours to reach a quasi-steady state \citep{FengandLiu2019,Reville_2020}. In contrast, for most MHD inner heliospheric models, the CFL-limited time steps are typically on the order of 10 minutes \citep{Detman2005,Hayashi2012}. As a result, coronal models demand significantly greater computational resources to remain synchronized with inner heliospheric models. Although empirical solar corona models \citep{Arge2003ImprovedMF,Yangzicai2018} offer higher efficiency, they discard important information and fail to deliver the necessary accuracy in forecasts \citep{Samara_2021}. Therefore, further efforts are needed to develop more efficient and accurate MHD coronal models. Implicit temporal discretization strategies can help overcome the limitations imposed by the CFL condition, thereby enhancing computational efficiency by allowing for larger time steps.

Recently, several successful attempts have been made to increase the efficiency of MHD coronal models by using implicit solvers \citep{brchnelova2023role,Feng_2021,Kuzma_2023,Liu_2023,Perri2018SimulationsOS,Perri_2022,Perri_2023,WANG2019181,WANG201967,Wang_2022,Wang2022_CJG,WangSubmitted}. In the implicit algorithm, convergence rate are improved by selecting a considerable time step. Although this approach may come with a loss in temporal accuracy, modeling the evolution and propagation of flux ropes has demonstrated that the implicit solar coronal model can maintain time accuracy while still outperforming the explicit MHD model in speed by selecting an appropriate time step size \citep{guo2023,Linan_2023}. Furthermore, \cite{WangSubmitted} employed the pseudo-time marching method to ensure temporal accuracy for CME simulations in the coronal region.
While these models have achieved the desired speedup, they remain constrained by a time-invariant magnetogram, which contrasts with the reality that the solar coronal structure evolves over time \citep{Owens2017}. This discrepancy leads to differences between simulation results and coronal observations \citep{Cash_2015,Reville_2020}.

To address the limitations of commonly used quasi-steady-state coronal models, which do not account for the evolution of coronal structures, many researchers have focused on developing time-evolving coronal models. These models, typically driven by hundreds of time-varying observed photospheric magnetograms, capture the evolution of coronal structures with higher fidelity \citep{Feng_2023,Yang2012,Yeates2018}. They may also enhance the realism of solar wind and CME modeling \citep{Lionello_2023}. However, it is too computationally expensive to perform time-evolving MHD coronal simulations \citep{Yeates2018}. One of the main reasons is that most state-of-the-art time-evolving coronal models \citep{Hayashi_2021,Hoeksema2020,Linker2024EGUGA,Lionello_2023,Mason_2023,Yang2012} still rely on explicit or semi-implicit approaches, where only certain source terms are treated implicitly while the time step remains constrained by the explicitly treated terms, leading to extremely low efficiency. As a result, real-time explicit or semi-implicit MHD coronal simulations typically require thousands of compute cores. For a more detailed description of time-evolving coronal models, see \cite{W_SubmittedCOCONUT}.

Furthermore, \cite{W_SubmittedCOCONUT} extended the quasi-steady-state COolfluid COroNal UnsTructured (COCONUT), a novel implicit MHD solar corona model based on the Computational Object-Oriented Libraries for Fluid Dynamics (COOLFluiD)\footnote{\url{https://github.com/andrealani/COOLFluiD.git}}, into a time-evolving coronal model. It is the first fully implicit time-evolving MHD coronal model, allowing the use of large time-step sizes exceeding the CFL condition. It can simulate the evolution of a full CR period in just 9 hours of computational time (with a time-step length of 10 minutes, 1,080 CPU cores, and approximately 1.5 million cells). However, it currently struggles with resolving low-$\beta$ issues. Given that $\left(\mathbf{B}+\epsilon~\mathbf{B}\right)^2-\mathbf{B}^2\equiv 2~\epsilon~\mathbf{B}^2+\epsilon^2~\mathbf{B}^2$, where $\epsilon~\mathbf{B}$ represent the magnetic field discretisation error, the magnetic pressure discretisation error can be comparable to thermal pressure in low-$\beta$ regions. This can lead to the emergence of non-physical negative thermal pressure when deriving thermal pressure from energy density. Reducing magnetic field discretization errors is crucial to mitigating these issues.

By solving decomposed MHD equations, where the magnetic field $\mathbf{B}$ is split into a time-independent potential magnetic field $\mathbf{B}_0$ and a time-dependent field $\mathbf{B}_1$ \cite[e.g.][]{FUCHS2010JCP,Guo2015,Powell1999,Tanaka1995}, researchers \cite[e.g.][]{Feng_2010,Feng_2021,Licaixia2018,WANG201967,Wang_2022,Wang2022_CJG,WangSubmitted} have reduced magnetic field discretization errors and significantly improved the numerical stability of quasi-steady-state MHD coronal modeling. However, as $\left|\mathbf{B}_1\right|$ increases in the time-evolving coronal simulations, discretization errors in $\mathbf{B}_1$ are still likely to result in non-physical negative thermal pressure and cause the code to break down. 
Therefore, \cite{Simulationsgriton2018} incorporated the rotation of the potential magnetic field $\mathbf{B}_0$ in simulations of planetary magnetospheres, treating $\mathbf{B}_0$ as a time-dependent field that co-rotates with the planet. This approach helps keep $\left|\mathbf{B}_1\right|$ small and improves the numerical stability of the MHD planetary magnetosphere code. 
Given that time-evolving coronal simulations involve not only differential rotation but also magnetic flux emergence and cancellation, we further introduce a temporally piecewise-constant variable in this paper to accommodate part of the non-potential field, ensuring that $\left|\mathbf{B}_1\right|$ remains consistently small throughout the simulations of the time-evolving corona.

Based on the above considerations, the paper is organized as follows. In Section \ref{sec:Governing equations and numerical methods} and Appendix \ref{DerivationofthedecomposedMHDequations}, we introduce the numerical formulation and implementations of the time-evolving MHD coronal model. The governing equations, the derivation of the extended magnetic field decomposition strategy, the processing of time-evolving boundary conditions, and the positivity-preserving (PP) measures applied to enhance the model's numerical stability are described in detail. In Section \ref{sec:Numerical Results}, we present the simulation results, including the evolution of the corona during CRs 2110 and 2111, as well as a comparison between the time-evolving simulation results and observational data. Finally, in Section \ref{sec:Conclusion}, we summarize the key features of the efficient and numerically stable fully implicit time-evolving coronal model and offer concluding remarks.

\section{Governing equations and numerical methods}\label{sec:Governing equations and numerical methods}
\subsection{The governing equations}\label{The governing equations}
This time-evolving MHD global coronal model is based on the time-accurate Solar Interplanetary Phenomena Implicit Finite Volume Method (SIP-IFVM) coronal model \citep{WangSubmitted}. Additionally, we incorporate the optically thin radiative loss and adjust the adiabatic index $\gamma$ from 1.05 to $\frac{5}{3}$ to better represent the adiabatic process. Furthermore, we develop an extended magnetic field decomposition strategy, the derivation of which is available in Appendix \ref{DerivationofthedecomposedMHDequations}, and employ it to enhance numerical stability of the time-evolving thermodynamic MHD coronal model. The governing equation is written as bellow

\begin{equation}\label{GoverEquationExtendedDecomposision}
\left\{
\begin{array}{l}
\frac{\partial \rho}{\partial t} + \nabla \cdot (\rho \, \mathbf{v}) = 0 \\
\frac{\partial (\rho \, \mathbf{v})}{\partial t} + \nabla \cdot \left[ \rho \, \mathbf{v \, v} + \left( p + \frac{\mathbf{B}^2}{2} - \frac{\mathbf{B}_{00}^2}{2} \right) \mathbf{I} - \mathbf{B \, B} + \mathbf{B}_{00} \, \mathbf{B}_{00} \right] \\
\quad = - \nabla \cdot \left( \mathbf{B}_1 + \mathbf{B}_{01} \right) \mathbf{B} - \frac{\rho G~M_s} {r^3}\mathbf{r}  + \mathbf{S}_m \\
\frac{\partial E_1}{\partial t} + \nabla \cdot \left[ \left( E_1 + p_{T1} + \mathbf{B}_1 \cdot \mathbf{B}_0 \right) \mathbf{v} - \mathbf{B} \left( \mathbf{v} \cdot \mathbf{B}_1 \right) \right] \\
\quad = - \nabla \cdot \left( \mathbf{B}_1 + \mathbf{B}_{01} \right) \left( \mathbf{v} \cdot \mathbf{B}_1 \right)
-\left(\mathbf{v} \cdot \nabla \right)\mathbf{B}_0 \cdot \mathbf{B}+\left(\mathbf{B} \cdot \nabla \right)\mathbf{B}_0 \cdot \mathbf{v}
- \rho \mathbf{v} \cdot \mathbf{r} \frac{G~M_s}{r^3} + S_E \\
\frac{\partial \mathbf{B}_1}{\partial t} + \nabla \cdot (\mathbf{v \, B} - \mathbf{B \, v}) = - \nabla \cdot \left( \mathbf{B}_1 + \mathbf{B}_{01} \right) \mathbf{v}
\end{array}
\right.
\end{equation}

In Eq. (\ref{GoverEquationExtendedDecomposision}), $\mathbf{B}=\left(B_x,B_y,B_z\right)^{T}$ and $\mathbf{v}=\left(u,v,w\right)^{T}$ denote the magnetic field and velocity in Cartesian coordinate system, $\rho$ is the plasma density, $\mathbf{B}_{00}=\left(B_{00x},B_{00y},B_{00z}\right)^{T}$ is a static potential field, $\mathbf{B}_{01}=\left(B_{01x},B_{01y},B_{01z}\right)^{T}$ is a temporally piecewise constant field and $\mathbf{B}_0=\mathbf{B}_{00}+\mathbf{B}_{01}$, $\mathbf{B}_1=\mathbf{B}-\mathbf{B}_{0}$, $E_1=\frac{p}{\gamma-1}+\frac{1}{2}\rho\mathbf{v}^{2}+\frac{1}{2}\mathbf{B}_1^{2}$ with the adiabatic index $\gamma=\frac{5}{3}$, and $p_{T1}=p+\frac{\mathbf{B}_1^2}{2}$.
During the simulations, as detailed in in Appendix \ref{DerivationofthedecomposedMHDequations}, when $\frac{p}{0.5 \, \mathbf{B}_1^2}$ falls below a specific threshold, we update $\mathbf{B}_{01}$, $\mathbf{B}_1$ and $E_1$ to $\mathbf{B}_{01}+\mathbf{B}_1$, $\mathbf{0}$ and $\frac{p}{\gamma-1}+\frac{1}{2}\rho\mathbf{v}^{2}$, respectively. In the definition of the magnetic field, a factor of $\frac{1}{\sqrt{\mu _0}}$ is absorbed with $\mu _0 = 4 \times 10^{-7} \pi ~ \rm H  ~ m^{-1}$ denoting the magnetic permeability.
Additionally, $G$ means the universal gravitational constant, $M_s$ means the mass of the Sun, and $G \, M_s=1.327927 \times 10^{20} ~\rm m^3~s^{-2}$.  The thermal pressure of the plasma is defined as $p= \Re \rho T$, where $\Re=1.653 \times 10^4 \rm m^2 s^{-2} K^{-1}$ denotes the gas constant, and $T$ is the temperature of the bulk plasma. Besides, $\mathbf{r}$ is the position vector and $r=\left|\mathbf{r}\right|$ refers to the heliocentric distance. $S_E=Q_e+\mathbf{v}\cdot\mathbf{S}_m+\nabla\cdot\mathbf{q} + Q_{rad}$ is the energy source term with the volumetric heating function $Q_e$ and $\mathbf{S}_m$ being used to mimic the effect of coronal heating and solar wind
acceleration, $Q_{rad}$ and $\nabla\cdot\mathbf{q}$ being used to mimic the radiation loss and thermal conduction.

By assuming the radiative losses to be optically thin \citep{Rosner1978,Zhou2021}, the radiative term $Q_{rad}$ is defined as below,
\begin{equation}\label{Radiationloss}
Q_{rad}=-n_e n_p \Lambda\left(T\right).\
\end{equation}
where the proton number density $n_p$ is assumed to be equal to the electric number density $n_e$ for the hydrogen plasma. Similar to \cite{WangSubmitted,van_der_Holst_2014}, the radiative cooling curve function $\Lambda\left(T\right)$ in this paper is derived from version 10 of CHIANTI \citep{Dere11997,CHIANTI10}, an atomic database for emission lines.
As did in \cite{Xia_2011,WangSubmitted}, we set $\Lambda\left(T\right)$ to zero when $T<2 \times 10^4\;$K, which means the plasma has become optically thick and is no longer fully ionized.
The heat flux $\mathbf{q}$ is defined in a Spitzer form or collisionless form mainly according to the radial distance \citep{Hollweg1978}:
\begin{equation}\label{heatflux}
\mathbf{q}=\left\{\begin{array}{c}
\xi T^{5 / 2}(\hat{\mathbf{b}} \cdot \nabla T) \hat{\mathbf{b}},  \text { if } 1 \leq r \leq 10 R_{s} \\
-\alpha n_{e} k_B T \mathbf{v},  \text { if } r>10 R_{s}
\end{array}\right.
\end{equation}
where $\hat{\mathbf{b}}=\frac{\mathbf{B}}{\left|\mathbf{B}\right|}$, $\xi =9 \times 10^{-12} \rm J \, m^{-1} \, s^{-1} \, K^{-\frac{7}{2}}$ \citep{Endeve_2003,Feng_2010}, the parameter $\alpha$ is set to $\frac{3}{2}$ \citep{Endeve_2003}, $n_{e}$ is the electron number density, and $k_B=1.380649 \times 10^{-23}\rm J \, K^{-1}$ is the Boltzmann constant. The Spitzer form of heat flux is defined along the magnetic field, whereas the collisionless form is defined along the velocity vector. To ensure a smooth transition between these two forms, we hybridize the Spitzer form $\mathbf{q}_{\rm Spitzer}$ and collisionless form $\mathbf{q}_{\rm Collisionless}$ based on the Alfv{\'e}nic Mach number within the domain $r \leq 10 R_{s}$:
\begin{equation}\label{heatfluxlt10Rs}
\mathbf{q}=\min\left(1, \frac{V_a}{\left|\mathbf{v}\right|}\right)\mathbf{q}_{\rm Spitzer}+\left(1-\min\left(1, \frac{V_a}{\left|\mathbf{v}\right|}\right)\right)\mathbf{q}_{\rm Collisionless},  \text { if } 1~R_s \leq r \leq 10~R_{s}
\end{equation}
where $V_a=\frac{\left|\mathbf{B}\right|}{\rho^{0.5}}$ is the Alfv{\'e}nic velocity.
Following the approach in \cite{Lionello_2008} and \cite{Mikic1999}, we apply a linear smoothing transition between Eq.~(\ref{heatfluxlt10Rs}) and $\mathbf{q}_{\rm Collisionless}$ over the domain of $7.5 \, R_s \leq r \leq 10 \, R_{s}$. In our time-evolving coronal simulation, we noticed that the plasma velocity can occasionally reach extraordinarily high values, exceeding 1000 $\rm km~s^{-1}$, in the vicinity of low-$\beta$ regions. This phenomenon can be alleviated by enhancing the heat conduction term. Therefore, we enhance the heat conduction efficiency in low-$\beta$ regions and ultimately adopt the following plasma $\beta$-dependent heat flux $\mathbf{q}^*$.
\begin{equation}\label{enhancedheatflux}
\mathbf{q}^*= \mathbf{q}   \left(1+\tanh\left(\frac{1}{\beta} \cdot \frac{1}{100}\right)\right)
\end{equation}

\subsection{Spatial discretization}\label{sec:Description of the spatialtemporal solver}
We adopt Godunov's method to advance cell-averaged solutions in time by solving a Riemann problem at each cell interface \citep{EINFELDT1991273,Godunov1959Adifference}. The computational domain is a spherical shell ranging from $1 R_s$ to $20 R_s$. The six-component composite mesh \citep{Feng_2010,Feng2020chap3,Wang_2022,WangSubmitted} with each identical component being a low-latitude spherical mesh confined by $\left(\frac{\pi}{4}-\delta _{\theta} \leq \theta \leq \frac{3 \pi}{4}+ \delta _{\theta}\right) \times \left(\frac{3 \phi}{4}-\delta _{\phi} \leq \phi \leq \frac{5 \pi}{4}+ \delta_{\phi}\right)$ and discretized to $96 \times 42 \times 42$ grid cells is adopted. Here $\delta _{\theta}$ and $\delta _{\phi}$ are adjustable parameters proportionally dependent on the grid spacing entailed for the minimum overlapping area.

As usual, the following discretized integral equations are obtained by integrating Eq.~(\ref{GoverEquationExtendedDecomposision}) over the hexahedral cell $i$ and applying Gauss's theorem to convert the volume integral of the flux divergence into a surface integral.
\begin{equation}\label{MHDequationdiscrization}
V_{i}\frac{\partial \mathbf{U}_{i}}{\partial t}+ \mathbf{R}_{i}= \mathbf{0},\
\end{equation}
where $\mathbf{R}_{i}=\oint_{\partial V_{i}}\mathbf{F}\cdot \mathbf{n} d \Gamma - V_{i}\mathbf{S}_{i}$ means the residual operator over cell $i$. Here $\oint_{\partial V_{i}}\mathbf{F}\cdot \mathbf{n} d \Gamma=\sum\limits_{j=1}^{6}\mathbf{F}_{ij}\left(\mathbf{U}_{L},\mathbf{U}_{R}\right)\cdot \mathbf{n}_{ij} \Gamma_{ij}$, $\mathbf{S}_i=\mathbf{S}_{{\rm Powell},i}+\mathbf{S}_{{\rm gra},i}+ \mathbf{S}_{{\rm heat},i}$ corresponds to the Godunov-Powell source terms, the gravitational force, and the heating and radiation loss source terms in cell $i$, respectively. $\mathbf{U}_{i}$ denotes the cell-averaged solution variables in cell $i$. As described in Appendix \ref{DerivationofthedecomposedMHDequations}, when $\min\limits_{\forall {\rm cell}_{i}}\frac{p_i}{0.5 \, \mathbf{B}_{1,i}^2}$ drops below a threshold, in this paper we adopt 10, we update $\mathbf{B}_{01,i}$ ,$\mathbf{B}_{1,i}=0$ and $E_{1,i}$ to $\mathbf{B}_{01,i}+\mathbf{B}_{1,i}$, $\mathbf{0}$ and $\frac{p_i}{\gamma-1}+\frac{1}{2}\rho_i\mathbf{v}_i^{2}$, respectively.

Additionally, in Eq.~(\ref{MHDequationdiscrization}), $V_{i}$ is the volume of cell $i$, $\Gamma_{ij}$ is the area of interface shared by cell $i$ and its neighbouring cell $j$, and $\mathbf{n}_{ij}$ is the unit normal vector of $\Gamma_{ij}$, pointing from cell $i$ to cell $j$.
Following the approach in \cite{Feng_2021} and \cite{Wang_2022}, the cell-averaged Powell source term is calculated as:
\begin{equation}\label{PowellSourceDiscritization}
\mathbf{S}_{{\rm Powell},i}=\frac{1}{V_i} \sum\limits_{j=1}^{6}\mathbf{T}^{-1}_{8ij}\mathbf{Q}\left(\mathbf{U}_{nL}\right)\left({B}_{1nL}+{B}_{01nL}-{B}_{1n,i}-{B}_{01n,i}-\eta\left({B}_{1nR}+{B}_{01nR}-{B}_{1nL}-{B}_{01nL} \right)\right) \Gamma_{ij}
\end{equation}
where $\eta=\frac{S_L}{S_R-S_L}$ \citep{Wu2019},
$
\mathbf{T}_{8ij}=\begin{pmatrix}1\quad \mathbf{0}\quad 0 \quad \mathbf{0}\\\mathbf{0} \quad \mathbf{T}_{ij}\quad \mathbf{0} \quad \mathbf{0}\\0 \quad \mathbf{0} \quad 1 \quad \mathbf{0} \\\mathbf{0} \quad \mathbf{0} \quad \mathbf{0} \quad \mathbf{T}_{ij}\end{pmatrix},\,
$
and $\mathbf{T}_{ij}=\begin{pmatrix} \mathbf{n}_{ij}, \mathbf{t}_{1ij} , \mathbf{t}_{2ij} \end{pmatrix}^T $ is a rotation matrix that transforms the $\left(x, y, z\right)$ coordinate system to the $\left(n, t_1, t_2\right)$ coordinate system. Here,  $\mathbf{t}_{1ij}$ and $\mathbf{t}_{2ij}$ are two unit orthogonal tangential vectors of cell face $\Gamma_{ij}$ \cite[e.g.,][and references therein]{Feng2020chapt2}, $S_L$ and $S_R$ are the velocities of two fast waves as defined in \cite{Wang_2022}, and $\mathbf{Q}\left(\mathbf{U}_{nL}\right)=\left(0, \, -\mathbf{T}_{ij} \cdot \mathbf{B}_{i,L}, \, -\mathbf{v}_{i,L}\cdot\mathbf{B}_{1i,L}, \, -\mathbf{T}_{ij} \cdot \mathbf{v}_{i,L}\right)^T$. The subscripts ``$_i$", ``$_{L}$" and ``$_{R}$" denote the corresponding variables at the centroid of cell $i$, and on $\Gamma_{ij}$ extrapolated from cell $i$ and from cell $j$, respectively. Besides, $\mathbf{S}_{{\rm gra},i}$, and $Q_{rad,i}$ included in $\mathbf{S}_{{\rm heat},i}$ are defined by evaluating their respective formulations using the corresponding variables at the centroid of cell $i$. The inviscid flux through the interface $\Gamma_{ij}$, described as $\mathbf{F}_{ij}\left(\mathbf{U}_{L},\mathbf{U}_{R}\right) \cdot \mathbf{n}_{ij}$, is computed using the positive-preserving (PP) HLL Riemann solver \citep{Feng_2021}, equipped with a self-adjustable dissipation term \citep{W_SubmittedCOCONUT}. The cell-averaged heat conduction term $\left(\nabla\cdot\mathbf{q}\right)_i$ is calculated following the Gauss's theorem, as described in \cite{Wang_2022}. The volume heating coefficients in $Q_{e,i}$ and $\mathbf{S}_{m,i}$ are precomputed from six-hourly updated magnetograms following the methods described in \cite{Feng_2021,Wang_2022,Wang2022_CJG}, and are stored as a time series and then linearly interpolated to the current time step during the time-evolving coronal simulations.

Following \cite{WangSubmitted}, the second-order positivity-preserving reconstruction method is used to reconstruct the piecewise polynomials of primitive variables
$\{\rho,u,v,w,p, T\}$ on the cell surface $\Gamma_{ij}$,
\begin{equation}\label{FlowfieldbyLSQ}
X_i(\mathbf{x})=\left.X\right|_i+\Psi_i\left.\nabla X\right|_i\cdot\left(\mathbf{x}-\mathbf{x}_i\right)
\end{equation}
where $X \in \{\rho, \, u, \, v, \, w, \, p, \, T\}$, $\left.X\right|_i$ is the corresponding variable at $\mathbf{x}_i$ (the centroid of cell $i$), $\left.\nabla X\right|_i=\left.\left(\frac{\partial X}{\partial x},\frac{\partial X}{\partial y},\frac{\partial X}{\partial z}\right)\right|_i$ is the derivative of $X$ at $\mathbf{x}_i$, and $\Psi_i$ is the limiter used to control spatial oscillation. Furthermore, a globally solenoidality-preserving (GSP) approach \citep{FengandLiu2019,Feng_2021} is employed to enhance the divergence-free constraint for the magnetic field. Considering that reducing the magnetic field discretization error is crucial for enhancing the positivity-preserving (PP) property of MHD models in low $\beta$ regions \citep{WangSubmitted}, and given that the MHD decomposition method introduced in this paper has improved the numerical stability of our code, we opt to abandon the limiter as follows in the calculation of the piecewise polynomial representations for $\mathbf{B}_{00}(\mathbf{x})$, $\mathbf{B}_{0}(\mathbf{x})$ and $\mathbf{B}_{1}(\mathbf{x})$ to minimize discretization error in the magnetic field.
\begin{equation}\label{B_GSP}
X_i(\mathbf{x})=\left.X\right|_i+\left.\nabla X\right|_i\cdot\left(\mathbf{x}-\mathbf{x}_i\right)
\end{equation}
where $X \in \{\mathbf{B}_{00x}, \, \mathbf{B}_{00y}, \, \mathbf{B}_{00z}, \, \mathbf{B}_{0x}, \, \mathbf{B}_{0y}, \, \mathbf{B}_{0z}, \,\mathbf{B}_{1x}, \, \mathbf{B}_{1y}, \, \mathbf{B}_{1z}\}$.

\subsection{Temporal integration}\label{sec:Temporal integration}
In this paper, we first apply the following backward Euler temporal integration to Eq.~(\ref{MHDequationdiscrization}) to perform a quasi-steady state coronal simulation constrained by one time-invariant magnetogram \citep{Feng_2021,Wang_2022}.
\begin{equation}\label{implicitbackwardEuler}
V_{i}\frac{\Delta \mathbf{U}^n_{i}}{\Delta t}+\mathbf{R}^{n+1}_{i}=\mathbf{0}.\
\end{equation}
The superscripts $^{`n'}$ and $^{`n+1'}$ denote the time level, $\Delta \mathbf{U}_i^{n}=\mathbf{U}_i^{n+1}-\mathbf{U}_i^{n}$, is the solution increment between the $n$-th and $(n+1)$-th time level, and $\Delta t = t^{n+1}-t^n$ is the time increment.
Subsequently, we evolve the magnetograms to carry out the time-evolving coronal simulation, enabling us to capture the temporal evolution of coronal structures. To improve the temporal accuracy for time-evolving simulations, we further employ the pseudo-time marching method \citep{WangSubmitted}, which introduced a pseudo time $\tau$ to Eq.~(\ref{implicitbackwardEuler}) and updated the solution during each physical time step $\Delta t$ by solving a steady-state problem on $\tau$, to solve Eq.~(\ref{implicitbackwardEuler}).
\begin{equation}\label{implicitbackwardEulerPseudotime}
V_{i}\frac{\Delta \mathbf{U}_{i}}{\Delta \tau}+\left(V_{i}\frac{\Delta \mathbf{U}^n_{i}}{\Delta t}+\mathbf{R}^{n+1}_{i}\right)=\mathbf{0}.\
\end{equation}
Here, $\Delta \tau$ is a pseudo time step and $\Delta \mathbf{U}_{i}$ is the solution increment during $\Delta \tau$.

Similar to the PP measure employed in \cite{W_SubmittedCOCONUT}, we make the following adjustment on the updated density and thermal pressure during each pseudo-time iteration in domain of $1 \, R_s \leq r \leq 2 \, R_{s}$.
\begin{equation}\label{RhoandPInner}
\left\{
\begin{array}{l}
\rho_i=\Upsilon_{\rho_i}\frac{\mathbf{B}_i^2}{V^2_{A,max}}+\left(1-\Upsilon_{\rho_i}\right)\rho_{o,i} \\
p_i=\Upsilon_{p_i}\frac{\mathbf{B}_{i}^2}{2}\beta_{\min}+\left(1-\Upsilon_{p_i}\right)p_{o,i}
\end{array}, \text { if } 1 \, R_s \leq r \leq 2 \, R_{s}
\right.
\end{equation}
In Eq.~(\ref{RhoandPInner}), $\Upsilon_{\rho_i}=0.5+0.5\cdot \tanh\left(\frac{V_{A,i}-V_{A,max}}{V_{fac}}\cdot \pi\right)$ with $V_{A,i}=\frac{\left|\mathbf{B}_i\right|}{\rho_{o,i}^{0.5}}$, $V_{A,max}=2\frac{\left|\mathbf{B}_{max}\right|}{\rho_{s}^{0.5}}$, and $V_{fac}=\frac{V_{A,max}}{1000}$. $\left|\mathbf{B}_{max}\right|$ represents the maximum magnetic field strength in the entire computational domain. Besides, $\Upsilon_{p_i}=0.5+0.5\cdot \tanh\left(\frac{\beta_{\min}-\frac{p_i}{0.5 \cdot \mathbf{B}_{i}^2+\epsilon}}{\beta_{fac}}\cdot \pi\right)$ with $\beta_{fac}= 10^{-7}$, $\beta_{\min}=10^{-4}$ and $\epsilon=10^{-12}$ are adopted in this paper.
Additionally, we constrain the plasma velocity in the range of $1 \, R_s \leq r \leq 1.1 \, R_{s}$ not to exceed the speed of sound $C_{s,i}=\left(\frac{\gamma\cdot p_i}{\rho_i}\right)^{0.5}$, as follows
\begin{equation}\label{vlimit}
\mathbf{v}_i=\mathbf{v}_{o,i}\cdot\min\left(\frac{C_{s,i}}{\left|\mathbf{v}_{o,i}\right|}\cdot\left(0.3+\tanh\left(\frac{r-R_s}{R_s}\cdot8.68\right)\right), \,1.0\right), \text { if } 1 \, R_s \leq r \leq 1.1 \, R_{s}.
\end{equation}
Here, the subscript ``$_{o}$" on $\rho$, $p$ and $\mathbf{v}$ denotes the density, thermal pressure and velocity updated during the pseudo-time iterations without adjustment.

During the time-evolving coronal simulation, the time-step length is gradually increased from an explicit time step to $\chi \cdot \tau_{flow}$ where $\chi$ is an adjustable parameter and $\tau_{flow}$ is a reference time length that is the same as defined in \cite{Feng_2021} and \cite{Wang_2022}. The time-step size can significantly affect the solution accuracy and computational efficiency for time-evolving simulations. A smaller time-step size increases the number of time steps required, demanding more computational resources. However, a particularly large time-step size can degrade temporal accuracy and may even result in program crashes \citep{WangSubmitted}. To make the time-step length achieve the necessary temporal accuracy, numerical stability, and high computational efficiency for time-evolving coronal simulations, we initially set $\chi=1$. It performs well for most of the time-evolving simulations.

Additionally, we noticed that during the long-term time-evolving simulations, a dramatic increase in the magnetograms, where the maximum magnetic field strength increases by more than 1.5 times, may occur between two adjacent magnetograms, potentially causing the code to crash. In this paper, this phenomenon occurred in the magnetograms at the 882-nd and 888-th hours of the approximately 1300-hour time-evolving simulation. Assuming such cases occur at a moment between $t_{m-1}$ and $t_{m}$, where the subscripts ``$_{m-1}$" and ``$_{m}$" correspond to the $\left(m-1\right)$-th and $m$-th magnetograms, we opt to employ a second-order Runge-Kutta method, with the intermediate states $\mathbf{U}^{(1)}$ and $\mathbf{U}^{(2)}$ computed using the backward Euler method \citep{Feng_2021,Wang_2022,Wang2022_CJG}, to solve Eq.~(\ref{MHDequationdiscrization}) over the time interval from $t_{m-2}$ to $t_{m+3}$.
\begin{equation}\label{2orderRK}
\begin{aligned}
&\mathbf{U}_i^{(1)}=\mathbf{U}_i^{n}-\Delta t \mathbf{R}_i^{(1)}\\
&\mathbf{U}_i^{(2)}=\mathbf{U}_i^{(1)}-\Delta t \mathbf{R}_i^{(2)}\\
&\mathbf{U}_i^{n+1}=\frac{1}{2}\left(\mathbf{U}_i^{n}+\mathbf{U}_i^{(2)}\right).
\end{aligned}
\end{equation}
During this time interval, the inner-boundary magnetic fields between $t_{m-2}$ and $t_{m+1}$ are linearly interpolated from the magnetograms at $t_{m-2}$ and $t_{m+1}$.

\subsection{Implementation of inner-boundary conditions}\label{sec:implementation of boundary conditions}
In this paper, we first perform a quasi-steady-state coronal simulation constrained by a time-invariant magnetogram \citep{Wang_2022}, and then evolve the magnetograms to drive the following time-evolving coronal simulation during CRs 2110 and 2111. Around this period, the time interval between two adjacent GONG-ADAPT magnetograms is 6 hour. And the original GONG-ADAPT magnetograms adopted in this paper are positioned in the co-rotating Carrington heliographic coordinate system. Therefore, we first rotated these magnetograms to the Heliocentric Inertial (HCI) coordinate system to match the inertial coordinate system. Subsequently, cubic Hermite interpolation is applied to these input magnetograms to derive the required inner-boundary magnetic fields for each time step during the time-evolving coronal simulations \citep{W_SubmittedCOCONUT}.

As did in \cite{WangSubmitted}, the inner-boundary conditions are specified at the inner boundary face, which coincides with the solar surface. Four Gaussian points are used on each inner-boundary face. The inner-boundary conditions at these Gaussian points, together with the solutions in the boundary cell and its seventeen neighboring cells that share at least a vertex with it, are utilized to construct the reconstruction formulation of primitive variables in the inner-boundary cells \citep{Feng_2021,WangSubmitted}. Take the $i$-th boundary cell, denoted as cell $BD,i$, which is a hexahedral cell consisting of one curved boundary surface and five planar faces \citep{Feng_2021,Wang_2022}, as an example:
\begin{equation}\label{Eq:FlowfieldIBD}
\begin{aligned}
X_{BD,i}(\mathbf{x})=&\left.X\right|_{BD,i}+\psi_{BD,i}\left.\left(\nabla X\right)\right|_{BD,i}\cdot\left(\mathbf{x}-\mathbf{x}_{BD,i}\right), \\ X \in &\{\rho,u,v,w,\mathbf{B}_1,\mathbf{B}_{00},\mathbf{B}_{0},p,T\},\
\end{aligned}
\end{equation}
The subscript ``$BD,i$" refers to the variable corresponding to the $i$-th boundary cell cell $BD,i$.

As the inner-boundary magnetic field evolves, a tangential component of the electric field, $\mathbf{E}_{BD,t}$, emerges at the inner boundary. According to the generalized Helmholtz theorem in classical electromagnetic theory, $\mathbf{E}_{BD,t}$ can be described as follows:
\begin{equation}\label{Helmholtz}
\mathbf{E}_{BD,t}=\nabla_t \, \Phi+\nabla_t\times\left(\Psi \, \mathbf{e}_r\right).
\end{equation}
where the subscript ``$_{BD,t}$" refers to the variable corresponding to the tangential component of a vector variable at the inner-boundary face. Here, $\Phi$ and $\Psi$ are arbitrary functions of position on the solar surface, and $\nabla_t$ represents the tangential derivative operator. The first term in the right-hand side of Eq.~(\ref{Helmholtz}) corresponds to the effects of the transverse magnetic field $\mathbf{B}_{BD,t}=\left(B_{BD, \theta}, B_{BD, \phi}\right)$, such as the transverse currents, while the second term represents variations in the radial magnetic field $B_{BD,r}$. Since the GONG-ADAPT magnetograms do not provide the transverse magnetic field, we consider only the first term in this study and defer the inclusion of the second term to future work. Combining this with Faraday's law,
$$\frac{\partial B_{BD,r}}{\partial t}=- \left(\nabla \, \times \mathbf{E}_{BD,t}\right)\cdot \mathbf{e}_r,$$
we obtain the following equation:
\begin{equation}\label{partialBpartialT}
\frac{\partial B_{BD,r}}{\partial t}=\nabla^2_t \, \Phi
\end{equation}
In this paper, Eq.~(\ref{partialBpartialT}) is solved on our six-component grid system using the 5-point method in a spherical coordinate. Considering that $\mathbf{E}_{BD,t}=-\left(\mathbf{v}_{BD} \times \mathbf{B}_{BD}\right)_t$, and given that only the radial component of the inner-boundary magnetic field is available at the inner-boundary face ($\mathbf{B}_{BD,t}=\mathbf{0}$), we derive the following equations for the tangential velocity components $\mathbf{v}_{BD,t}=\left(v_{BD,\theta}, v_{BD,\phi}\right)$:
\begin{equation}\label{Vthetaph}
v_{BD,\theta}=\frac{E_{BD,\phi}}{B_{BD,r}}, \, v_{BD,\phi}=-\frac{E_{BD,\theta}}{B_{BD,r}}
\end{equation}
where the subscripts ``$_\theta$" and ``$_\phi$" indicate the respective vector components along the $\theta$ and $\phi$ directions.
Additionally, following \cite{Lionello_2023}, we apply the following adjustment to Eq.~(\ref{Vthetaph}) to regularize the flow near the boundary polarity inversion line, where $B_{BD,r}=0$:
\begin{equation}\label{Vthetaphregulated}
\mathbf{v}_{BC,t}=\frac{\left(\mathbf{E}_{BC}\times \mathbf{B}_{BC}\right)_t}{\mathbf{B}_{BC}^2+D\left|\mathbf{E}_{BC}\right|\cdot\left|\mathbf{B}_{BC}\right|\big/C_s}
\end{equation}
This means that the magnitude of the tangential velocity will converge to $\frac{C_s}{D}$, where $D=3$ is adopted in this paper and $C_s$ is the local sound speed calculated from the solution in the cell immediately adjacent to the inner-boundary cell in the radial direction.

During the simulations, the boundary conditions for thermal pressure $p_{BD}$, plasma density $\rho_{BD}$, temperature $T_{BD}$, and radial velocity $v_{BD,r}$ is classified into two cases based on the radial velocity $v_r$ in the cell immediately adjacent to the inner-boundary cell in the radial direction \citep{Groth2000,Feng_2021,Wang_2022,Wang2022_CJG,WangSubmitted}.
When $v_r\geq0$:
The thermal pressure and plasma density at the inner-boundary face are set as $p_{BD}=\frac{1}{\gamma}$ and $\rho_{BD}=1$ in code unit. Following \cite{brchnelova2023assessing,W_SubmittedCOCONUT}, the inner-boundary density $\rho_{BD}$  is further adjusted based on the local Alfv{\'e}nic velocity, as in Eq.~(\ref{RhoandPInner}), to enhance the positivity-preserving property of the coronal model. Additionally, the inner-boundary temperature is calculated as $T_{BD}=\frac{\gamma \cdot p_{BD}}{\rho_{BD}}$, and the radial velocity is constrained as $\frac{\partial v_{BD,r}}{\partial r}=0$.
When $v_r \leq 0$:
$\frac{\partial \rho_{BD}}{\partial r}=0$, $\frac{\partial p_{BD}}{\partial r}=0$, $\frac{\partial T_{BD}}{\partial r}=0$ and $v_{BD,r}=0$ are applied.

Additionally, the variables at the centroid of inner-boundary cells are required in calculation of Eq.~(\ref{Eq:FlowfieldIBD}). In this paper, the magnetic field is generated from the potential field (PF) solver with a 15-order spherical harmonic expansion. The tangential velocity is computed as an average of the values at four Gaussian points and at the centroid of the neighboring cell immediately adjacent to the inner-boundary cell in the radial direction. The radial velocity, thermal pressure, plasma density, and temperature are derived using Parker's one-dimensional hydrodynamic isothermal solar wind solution \citep{parker1963}. Besides, the thermal pressure and plasma density are adjusted using Eq.~(\ref{RhoandPInner}).

\section{Numerical results}\label{sec:Numerical Results}
\begin{figure}[htpb]
\begin{center}
  \vspace*{0.01\textwidth}
    \includegraphics[width=0.8\linewidth,trim=1 1 1 1, clip]{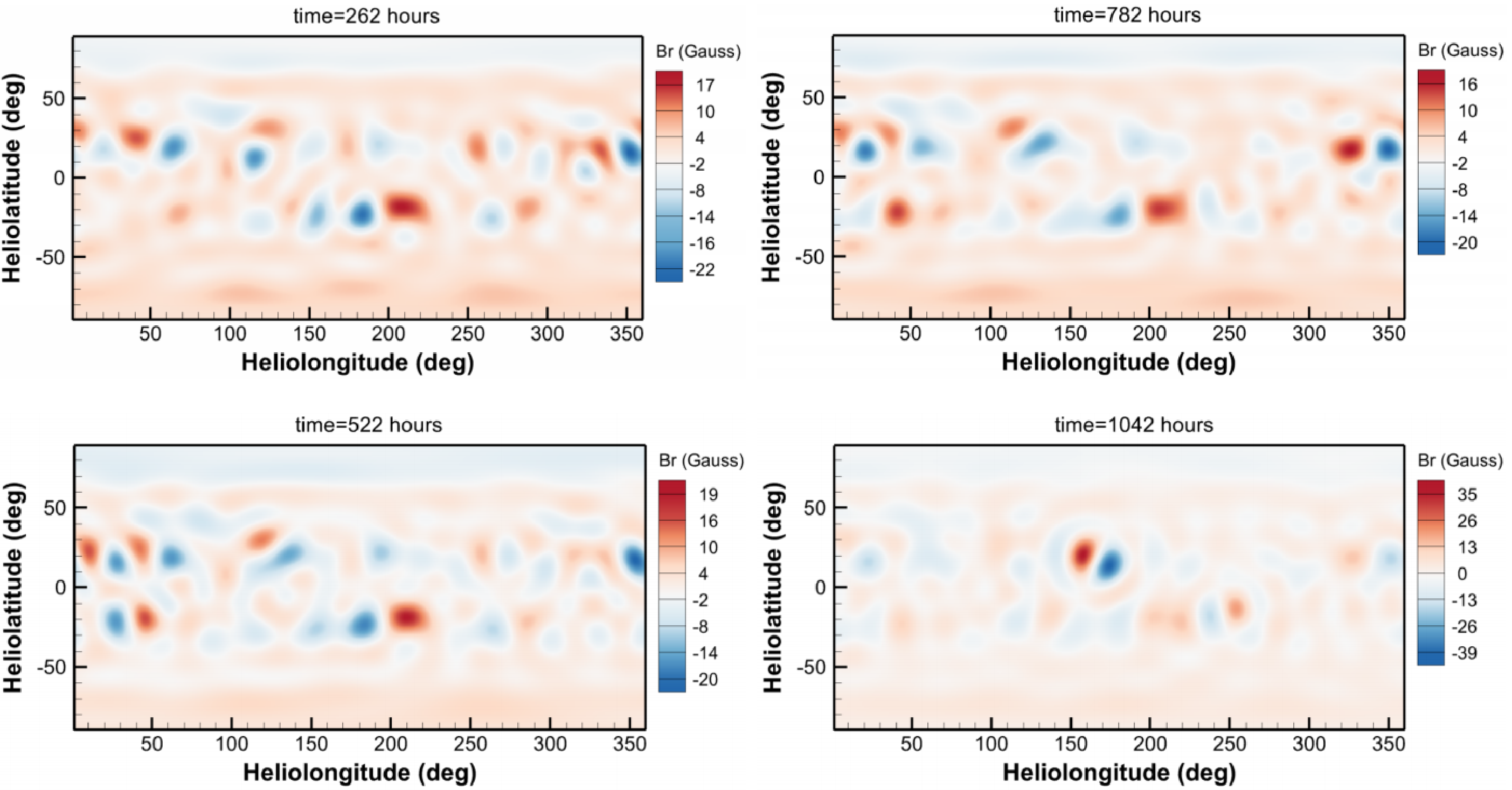}
\end{center}
\caption{Distribution of the radial magnetic field used as the inner boundary condition at the solar surface, shown in a co-rotating coordinate system.}\label{magneticfieldatsolarsurface}
\end{figure}

In this section, the time-evolving SIP-IFVM coronal model, equipped with the extended magnetic field decomposition strategy, is employed to simulate the evolution of coronal structures during Carrington Rotations (CRs) 2110 and 2111. This period, spanning from May 9 to July 3, 2011, corresponds to the solar maximum of Solar Cycle 24. Approximately 220 GONG-ADAPT magnetograms\footnote{\url{https://gong.nso.edu/adapt/maps/gong/2011/}}, with the first realization of the 12-member ensemble \citep{Perri_2023} adopted and updated at a 6-hour cadence, are used to drive these simulations in an inertial coordinate system spanning from the solar surface to 20 $R_s$. Fig.~\ref{magneticfieldatsolarsurface} displays the inner-boundary magnetic field at various moments. For convenience of comparatione, these magnetic field distributions are illustrated in a co-rotating coordinate system. It reveals that the inner-boundary magnetic field varies more significantly at low and middle latitudes, while the regions beyond $70^\circ$ in both the north and south poles are predominantly occupied by magnetic fields directed inward and outward from the Sun. Additionally, the magnetic field strength approximately doubles during these two CRs. During the time-evolving simulations, the magnetic field near the solar surface sometimes exceeds 40 Gauss with the plasma beta reaching a minimum value of around $10^{-3}$. The entire simulation consists of 23040 time steps, with the average time step being approximately 3.4 minutes.

\begin{figure}[htpb]
\begin{center}
  \vspace*{0.01\textwidth}
    \includegraphics[width=0.8\linewidth,trim=1 1 1 1, clip]{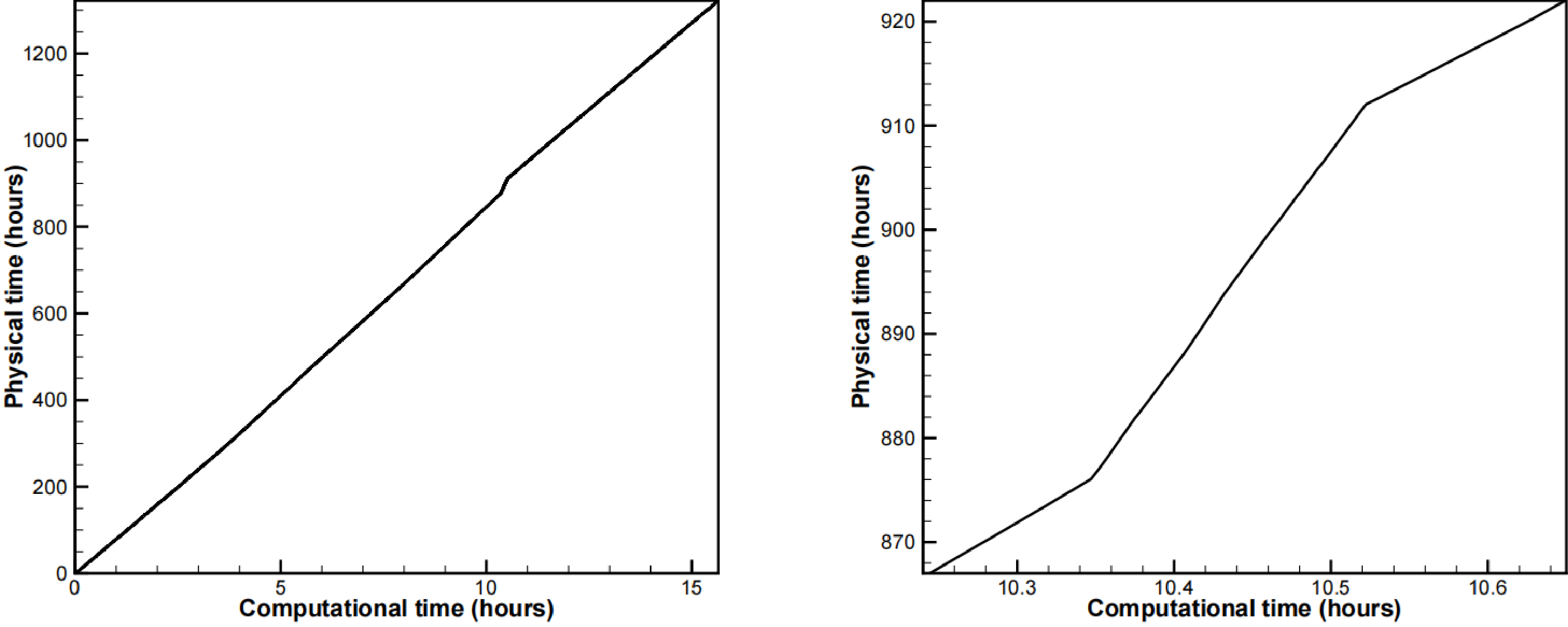}
\end{center}
\caption{Evolution of physical time versus computational time in the time-evolving coronal simulation for CRs 2110 and 2111 (left) and a close-up view of the profile between 10.24 and 10.65 hours of computational time (right).}\label{PhysicalTimeVersusComputationalTime}
\end{figure}

All calculations in this paper are performed on the WICE cluster, part of the Tier-2 supercomputer infrastructure of the Vlaams Supercomputer Centrum (VSC)\footnote{\url{https://www.vscentrum.be/}}. By utilizing 192 CPU cores, the wall-clock time for the simulation of these two solar-maximum CR periods is approximately 16 hours. The evolution of physical time along the computational time is illustrated in Fig.~\ref{PhysicalTimeVersusComputationalTime}. It demonstrates that the evolution of coronal structures over 1322 hours of physical time is completed within just 16 hours of computational time, showcasing that our model operates over 80 times faster than real-time coronal evolution using only 192 CPU cores. Additionally, the close-up view of the profile between 876 and 912 hours of physical time, which is around the event of a dramatic increase in the magnetograms described in Subsection \ref{sec:Temporal integration}, is completed within 0.18 computational hours, demonstrating an efficiency approximately 1.5 times faster than during other periods. While this strategy may miss some high-frequency phenomena during this period and reduce temporal accuracy, it effectively prevents code crashes around the dramatic increase in the magnetograms.

In Subsection \ref{OpenField}, we display the distribution of open- and closed-magnetic field regions and compare the simulated extreme ultraviolet (EUV) images with observations. In Subsection \ref{CoronalNeartheSun}, we present the simulated and observed white-light polarized brightness (pB) images. We also show the one- and two-dimensional (1 D and 2 D) timing diagrams of some simulated parameters at 3 $R_s$. In Subsection \ref{Insitu}, we demonstrate the timing diagrams of some selected parameters monitored by some virtual satellites and map the radial velocity and magnetic field observed at 1 AU back to 20 $R_s$ for comparison. We also illustrate the 2 D timing diagrams of the simulated radial velocity and plasma density at 20 $R_s$.

\subsection{Distributions of the open and closed-magnetic field regions}\label{OpenField}
Coronal holes (CHs) are dark regions observed in extreme ultraviolet (EUV) and soft X-ray channels, typically associated with low plasma density and magnetic field lines that are open to interplanetary space. CH distributions vary across different phases of solar activity and are among the most prominent features of the solar corona \citep{FengMa2015,Feng_2017,FengandLiu2019,Frazin2007,Hayes2001,Linker1999JGR,Petrie2011SoPh}. Generally, three types of CHs can be identified in EUV and soft X-ray images of the solar corona. Polar CHs are located at the solar poles and often extend to lower latitudes, occasionally crossing the solar equator. Isolated CHs, commonly observed near solar maxima, are detached from polar CHs and scattered across low and mid-latitudes. Transient CHs are associated with solar eruptive events, such as coronal mass ejections, solar flares, and eruptive prominences.

\begin{figure}[htpb]
\begin{center}
  \vspace*{0.01\textwidth}
    \includegraphics[width=0.8\linewidth,trim=1 1 1 1, clip]{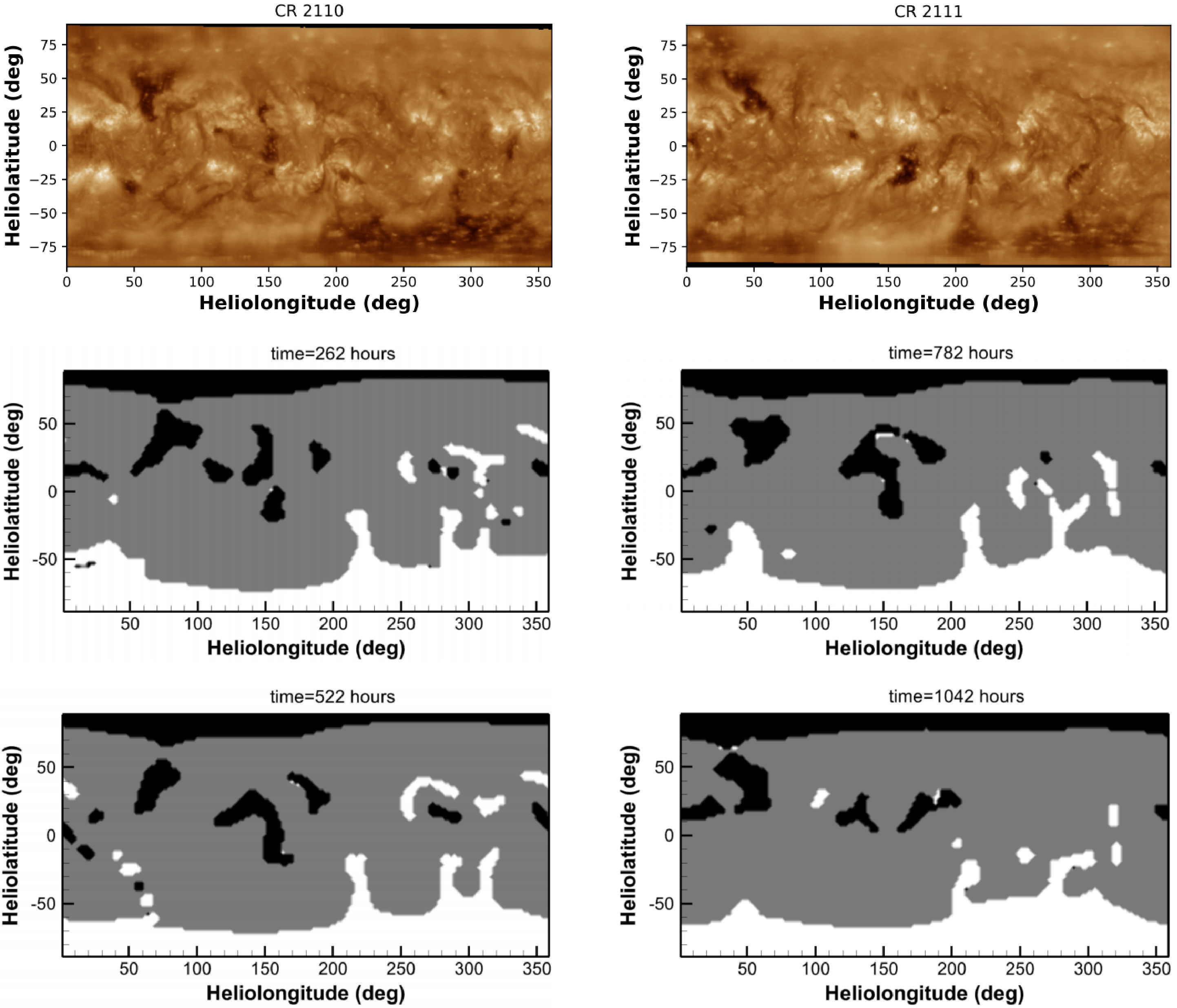}
\end{center}
\caption{Synoptic maps of the EUV observations from the 193 {\AA} channel of AIA on board SDO (top) for CRs 2110 (left) and 2111 (right), alongside the distributions of open- and closed-field regions modelled by the time-evolving SIP-IFVM coronal model (middle and bottom), all shown in a co-rotating coordinate system. In the middle and bottom panels, the white and black patches represent open-field regions with magnetic field lines pointing outward and inward relative to the Sun, respectively, while the gray patches indicate closed-field regions.}\label{CHdistribution}
\end{figure}

In Fig.~\ref{CHdistribution}, we derive the simulated open- and closed-field regions by tracing the magnetic field lines from 2.5 $R_s$ to the solar surface(middle and bottom) and compare them with the synoptic maps of the observations (top) from the 193 {\AA} channel of Atmospheric Imaging Assembly (AIA 193 {\AA}) telescope \citep{Lemen2012} onboard the Solar Dynamics Observatory spaceship \citep{Pesnell_2012SoPh}\footnote{\url{https://sdo.gsfc.nasa.gov/data/synoptic/}}. These images are illustrated in a co-rotating coordinate system. It reveals that the simulations aptly reproduce the observed southern polar CH with the latitudes $70^{\circ}$ poleward. The leading CHs around $220^{\circ}$ and $280^{\circ}$ in longitude, extending from the southern pole to near the solar equator, as well as the isolated CH centred at $(\theta_{\rm lat},\phi_{\rm long})=(40^{\circ} {\rm N}, 60^{\circ})$ and detached from the northern polar CH, are also well captured. Here, ``$\theta_{\rm lat}$" represents the heliographic latitude, and ``$\phi_{\rm long}$" denotes the Carrington longitude. Additionally, the decreasing trend of scattered isolated CHs in the low-latitude domain from CR 2110 to 2111 is also captured in the simulation results. Notably, the isolated CH centered around $(\theta_{\rm lat},\phi_{\rm long})=(0^{\circ}, 150^{\circ})$ rapidly transformed from an upside-down hook shape at 782 hours to a whirlwind-like structure above the solar equator at 1042 hours. Together with Fig.~\ref{magneticfieldatsolarsurface}, this rapid change can be attributed primarily to the emergence of a dipole centered at $(\theta_{\rm lat},\phi_{\rm long})=(10^{\circ} {\rm N}, 160^{\circ})$.

\begin{figure}[htpb]
\begin{center}
  \vspace*{0.01\textwidth}
    \includegraphics[width=0.8\linewidth,trim=1 1 1 1, clip]{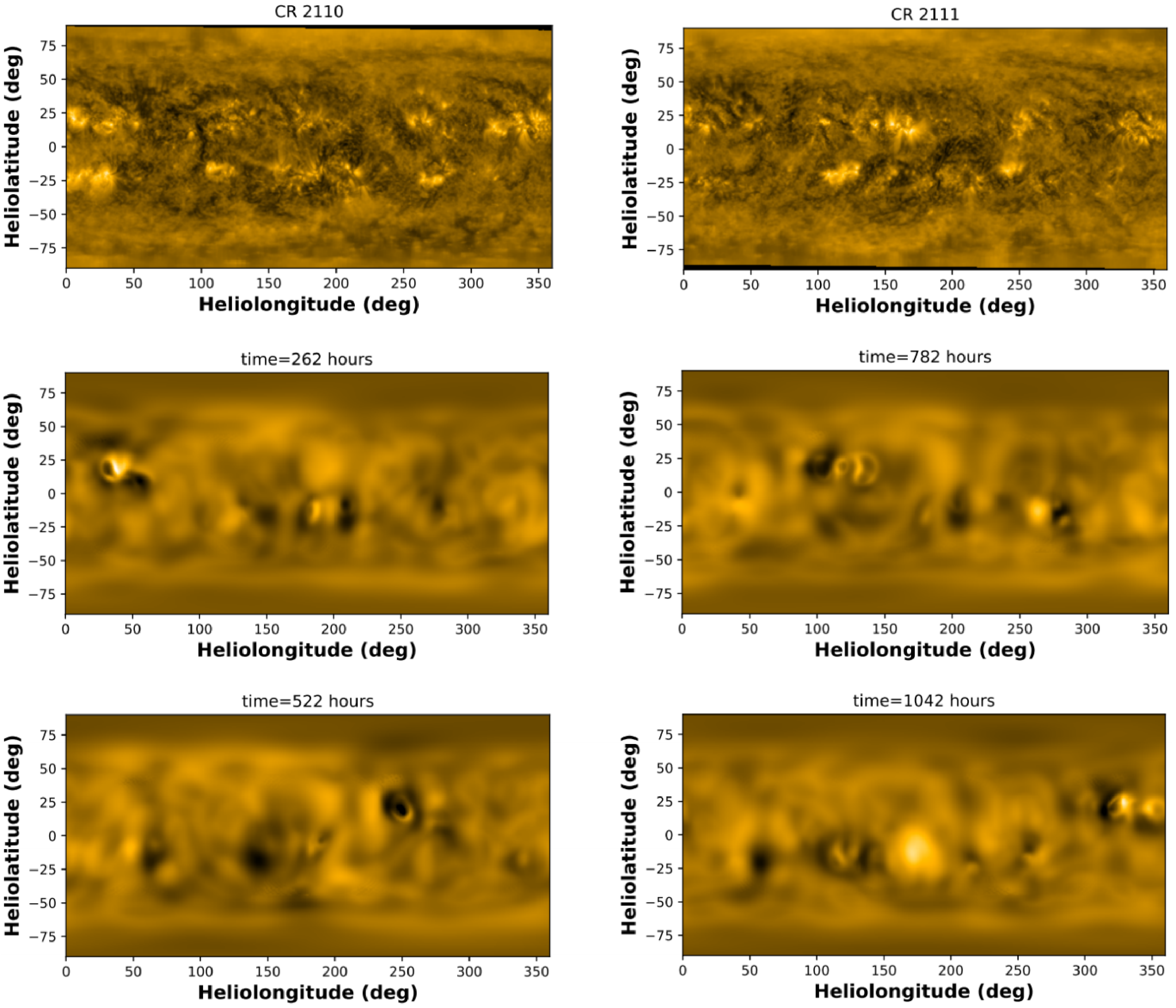}
\end{center}
\caption{Synoptic maps of the EUV observations from the 171 {\AA} channel of AIA on board SDO (top) for CRs 2110 (left) and 2111 (right), alongside the EUV images derived from the time-evolving simulation results at different moments (middle and bottom).}\label{EUV171distribution}
\end{figure}

In Fig.~\ref{EUV171distribution}, we present the AIA 171 {\AA} EUV observations (top) along with the simulated 171 {\AA} EUV images (middle and bottom) in a co-rotating coordinate system. The simulated EUV images are derived by integrating along the radial direction from 1.02 $R_s$ to 1.5 $R_s$. The emission in each point $\mathbf{x}$ of our computational domain is calculated via 
$$I_{\lambda}(\mathbf{x}) = G_{\lambda}(T) n_e^2(\mathbf{x}),$$
where the wavelength $\lambda=171 {\AA}$ and $G_{\lambda}$ is the temperature dependent response function for the $171 {\AA}$ waveband, which is obtained from the CHANTI atomic database \citep{Dere11997,CHIANTI10}. Meanwhile, the synoptic AIA 171 {\AA} EUV observation images are generated by concatenating a series of meridional strips extracted from full-disk images over the duration of an entire CR \citep{Hamada2018}.
The 171 {\AA} EUV images primarily show the solar corona at temperatures of approximately from 0.8 $\sim$ 1.0 million K (M K). It is mainly sensitive to emission from Fe IX (iron-9) ions, which are abundant in the quiet corona and upper transition region, highlighting coronal loops and other magnetic structures in the low corona regions.

Since the lifetime of coronal loops ranges from minutes and hours to weeks, and our model resolution does not resolve the transition region, the simulation results do not accurately capture the bright structures observed in the AIA 171 {\AA} EUV images. However, the dark regions, spanning longitudes $250^{\circ}$ to $360^{\circ}$ between $50^{\circ} {\rm S}$ and $75^{\circ} {\rm N}$, and $60^{\circ}$ to $200^{\circ}$ between $50^{\circ} {\rm S}$ and $50^{\circ} {\rm N}$, are basically captured. The observed bright structures centered at $(\theta_{\rm lat},\phi_{\rm long})=(20^{\circ} {\rm N}, 40^{\circ})$ and $(20^{\circ} {\rm S}, 190^{\circ})$ for CR 2110 are reproduced in the simulation at 262 hours. The observed bright structure centered at $(\theta_{\rm lat},\phi_{\rm long})=(25^{\circ} {\rm N}, 270^{\circ})$, accompanied by a dark region on its left side, is captured by the simulation at 522 hours. For CR 2111, the simulation at 782 hours approximately reproduces the observed bright structures centered at $(\theta_{\rm lat},\phi_{\rm long})=(20^{\circ} {\rm N}, 150^{\circ})$ and $(20^{\circ} {\rm S}, 250^{\circ})$. Additionally, the bright structure centered at $(\theta_{\rm lat},\phi_{\rm long})=(20^{\circ} {\rm N}, 340^{\circ})$ is well captured at 1042 hours. The extremely bright structure centered at $(\theta_{\rm lat},\phi_{\rm long})=(15^{\circ} {\rm S}, 170^{\circ})$ at 1042 hours may be attributed to the transition of the open-field region in this area around 782 hours to a closed-field configuration at 1042 hours.

emergence of the dipole centered at $(\theta_{\rm lat},\phi_{\rm long})=(10^{\circ} {\rm N}, 160^{\circ})$, as illustrated in Fig.~\ref{magneticfieldatsolarsurface}.

\subsection{Coronal structures near the Sun}\label{CoronalNeartheSun}
White-light polarized brightness (pB) images can reveal various large-scale coronal structures. In these pB images, bright regions correspond to high-density coronal structures, such as bipolar streamers and pseudo-streamers, while dark regions indicate low-density structures, such as coronal holes\citep{FengMa2015,Feng_2017,FengandLiu2019,Frazin2007,Hayes2001,Linker1999JGR,Petrie2011SoPh,Feng2020book}. Bipolar streamers separate CHs with opposite magnetic polarities, while pseudo-streamers separate CHs of the same polarity. Additionally, bipolar streamers extend outward several solar radii from the Sun, forming a cusp-like structure as they are drawn into a current sheet above the helmet streamer \citep{Abbo2015,Feng_2017,FengandLiu2019,Riley2011,Wang_2007}.

\begin{figure}[htpb]
\begin{center}
  \vspace*{0.01\textwidth}
    \includegraphics[width=\linewidth,trim=1 1 1 1, clip]{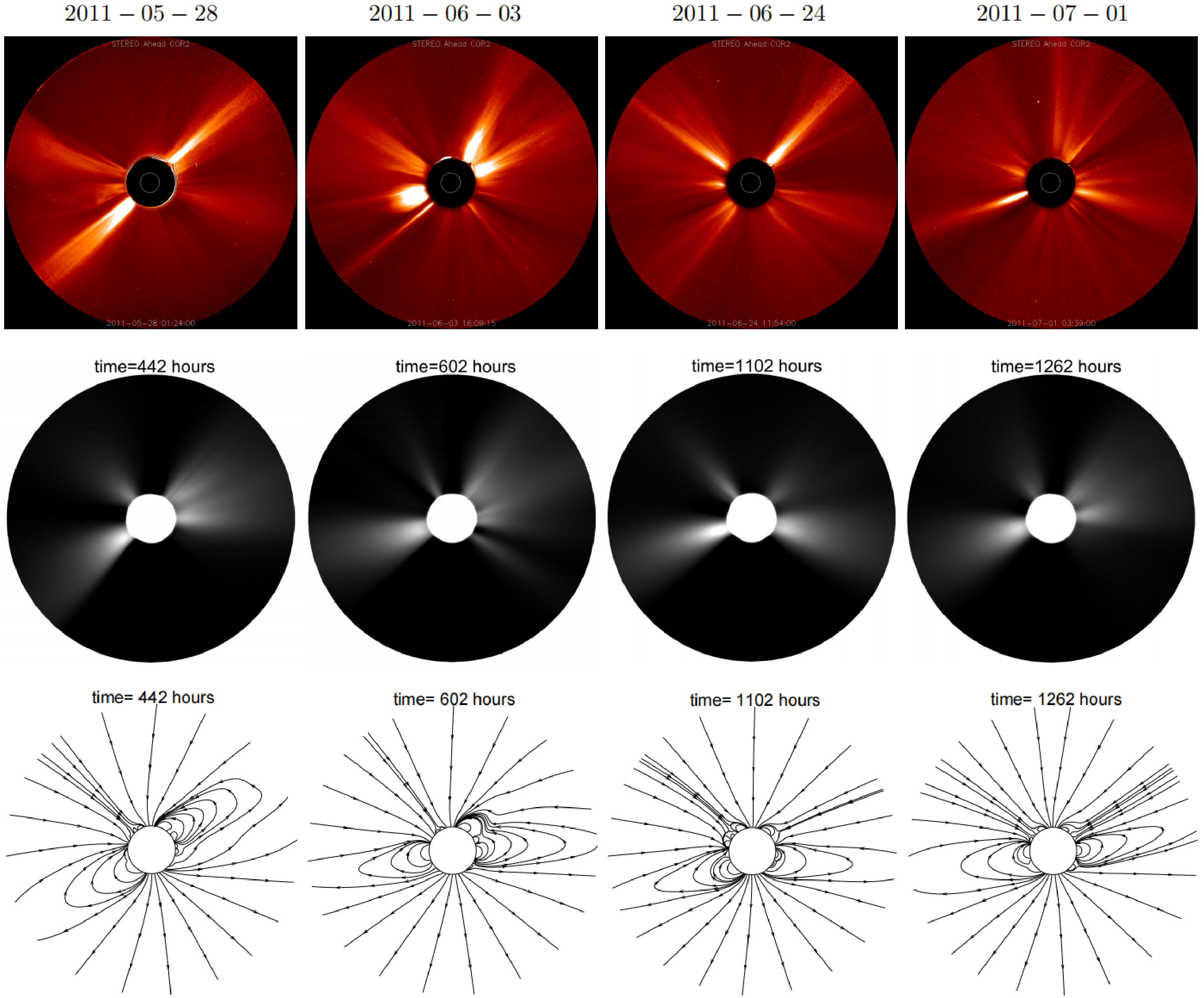}
\end{center}
\caption{White-light pB images observed by COR2/STEREO-A (top), ranging from 2.5 to 15 $R_s$, alongside corresponding pB images synthesised from simulation results ranging from 2.5 to 15 $R_s$ (middle) and 2 D distributions of some selected simulated magnetic field lines from 1 to 6 $R_s$ (bottom). These images are shown on the meridional plane in the STEREO-A view.}\label{pBsky}
\end{figure}

In Fig. \ref{pBsky}, we compare white-light pB images from the innermost coronagraph of the Sun Earth Connection Coronal and Heliospheric Investigation (SECCHI) instrument suite (top) onboard the Solar Terrestrial Relations Observatory Ahead (STEREO-A) spacecraft \footnote{\url{https://stereo-ssc.nascom.nasa.gov/browse/}}  \citep{Howard2008,Kaiser2008TheSM,Thompson2003COR1IC,Thompson2008} with those synthesized from the simulation results ranging from 2.5 to 15 $R_s$ (middle). The 2 D distributions of some selected simulated magnetic field lines on the meridional plane in the STEREO-A view, ranging from 1 to 6 $R_s$, are also illustrated (bottom). This comparison indicates that the simulated magnetic field lines and bright structures are generally consistent with the observed bipolar and pseudo-streamers, although some discrepancies exist in the width and position of these bright structures. The mismatch may come from the imperfect measurements of photospheric magnetic fields, possibly presence of the solar disturbances during the two CRs that are not considered in the MHD modeling.

At 442 hours in the time-evolving coronal simulation, both the simulation and observations exhibit two bright structures centered around $44^{\circ} {\rm S}$ and $46^{\circ} {\rm N}$ at the east and west limbs, respectively. The magnetic field lines indicate that these bright structures are formed by bipolar streamers. However, in the simulation, the observed bright structures centered around $10^{\circ} {\rm N}$ and $20^{\circ} {\rm S}$ at the east and west limbs are shifted northward by approximately $40^{\circ}$ and $20^{\circ}$, respectively. At 602 hours, the three bright structures centered around $22^{\circ} {\rm S}$ at the east limb and $57^{\circ} {\rm N}$ and $7^{\circ} {\rm N}$ at the west limb are well reproduced in the simulation, formed by bipolar streamers. However, the pseudo streamers located around $34^{\circ} {\rm S}$ and $61^{\circ} {\rm N}$ at the west and east limbs appear $10^{\circ}$ and $20^{\circ}$ farther north than in the observations. At 1102 hours, the simulated bipolar streamer around $12^{\circ} {\rm S}$ at the west limb successfully captures the observed bright structure, while the simulated bipolar streamer around $22^{\circ} {\rm S}$ at the east limb is shifted $20^{\circ}$ northward compared to observations. Additionally, the pseudo streamers around $48^{\circ} {\rm N}$ at both limbs align well with the observed bright structures. At 1262 hours, the simulated bipolar streamer around $20^{\circ} {\rm S}$ at the east limb and $6^{\circ} {\rm N}$ at the west limb effectively reproduce the observed structures. The pseudo streamers around $45^{\circ} {\rm N}$ and $59^{\circ} {\rm N}$ at the west and east limbs also closely match the observations. However, the observed bright structure around $18^{\circ} {\rm S}$ is not present in the simulation results, despite the existence of a pseudo streamer in the magnetic field structure.

\begin{figure}[htpb]
\begin{center}
  \vspace*{0.01\textwidth}
    \includegraphics[width=0.8\linewidth,trim=1 1 1 1, clip]{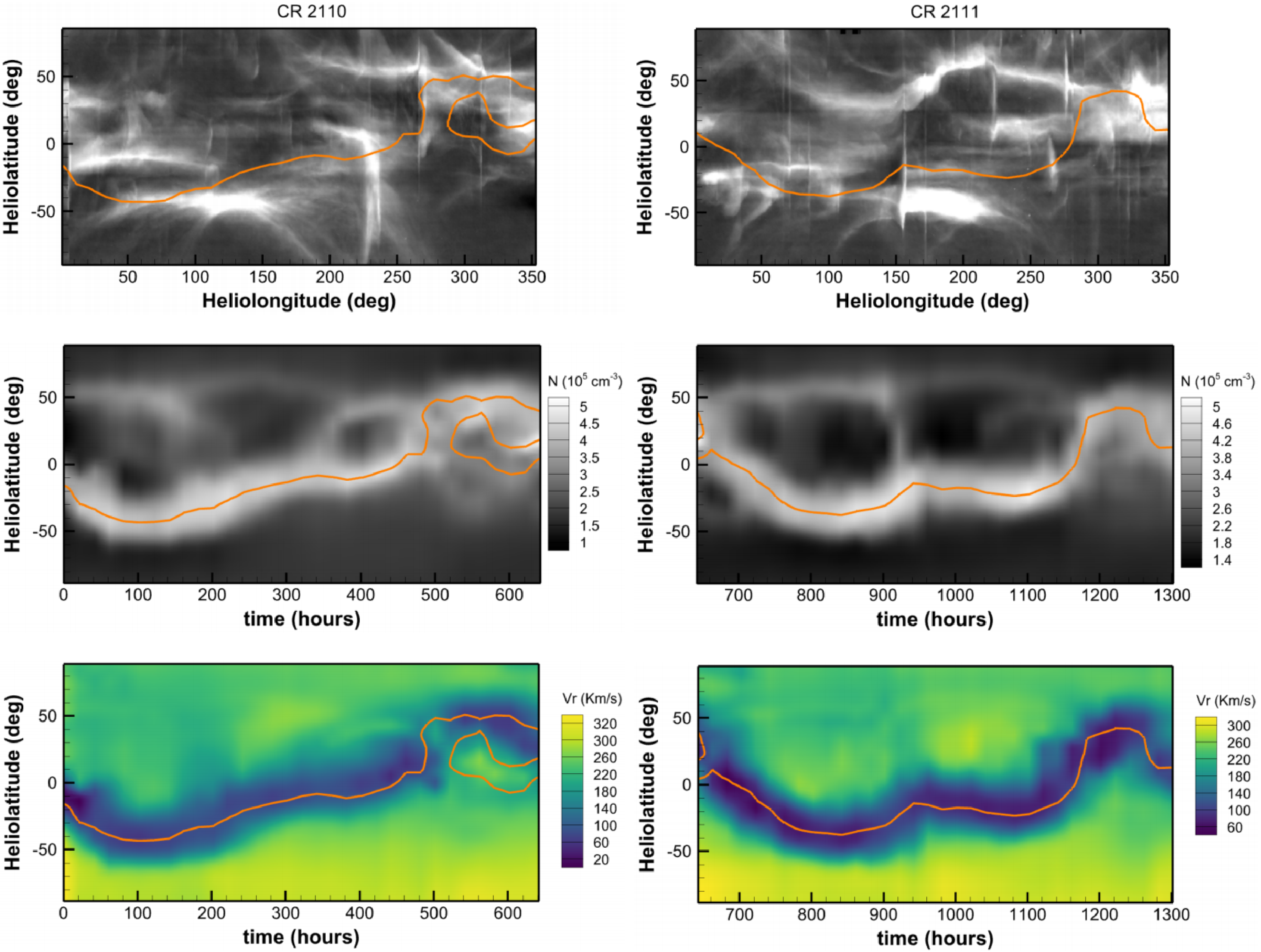}
\end{center}
\caption{Synoptic maps of east-limb white-light pB images at 3 $R_s$ observed by the LASCO C2 instrument onboard SOHO satellite (top) for CRs 2110 (left) and 2111 (right), alongside the timing diagrams of simulated plasma number density ($\rm 10^5 ~ cm^{-3}$) and radial velocity $V_r$ ($\rm km~s^{-1}$ at 3 $R_s$ at the east-limb longitude of the Sun in the Earth view.}\label{pBat3Rs}
\end{figure}

Additionally, we compare the pB observations at $3 R_s$ with the modeled results in Fig. \ref{pBat3Rs}. The top panels illustrate synoptic maps of the east-limb observations from the Large Angle and Spectrometric Coronagraph C2 (LASCO-C2) \citep{Brueckner1995} onboard the Solar and Heliospheric Observatory (SOHO) \footnote{\url{https://sdo.gsfc.nasa.gov/data/synoptic/}} for CRs 2110 (left) and 2111 (right), with the bright structures representing distributions of the high-density coronal structures. The middle and bottom panels display the 2 D timing diagrams of simulated plasma number density ($\rm 10^5 ~ cm^{-3}$) and radial velocity $V_r$ ($\rm km~s^{-1}$). These 2 D timing diagrams are synthesized from a series of time-evolving simulation results with a cadence of one result per 20 hours, the radial basic function (RBF) interpolation method \citep{Wang_2022} is applied to interpolate the variables to the east-limb longitude of the Sun in the Earth view. The orange solid lines overlaid on these counter denote the magnetic neutral lines (MNLs) modeled by the time-evolving coronal model.

The comparison in Fig. \ref{pBat3Rs} demonstrates that the bright structures in the simulated plasma density are generally consistent with those in the observed pB observations. The simulated high-density, low-speed flows are primarily distributed around the MNLs. Additionally, it is noted that the northernmost bright structures in the simulation for CR 2110 extend approximately $20^{\circ}$ farther north during the first 320 hours. Meanwhile, for CR 2111, the bright structures during the first 270 hours of the simulation changed from a trapezoidal cavity to irregular solid triangles in the observations. These mismatches may be attributed to the fact that the magnetic field at different longitudes in each synoptic magnetographs is observed at different times, as well as the imperfect measurements of the photospheric magnetic fields in the polar regions.

\begin{figure}[htpb]
\begin{center}
  \vspace*{0.01\textwidth}
    \includegraphics[width=\linewidth,trim=1 1 1 1, clip]{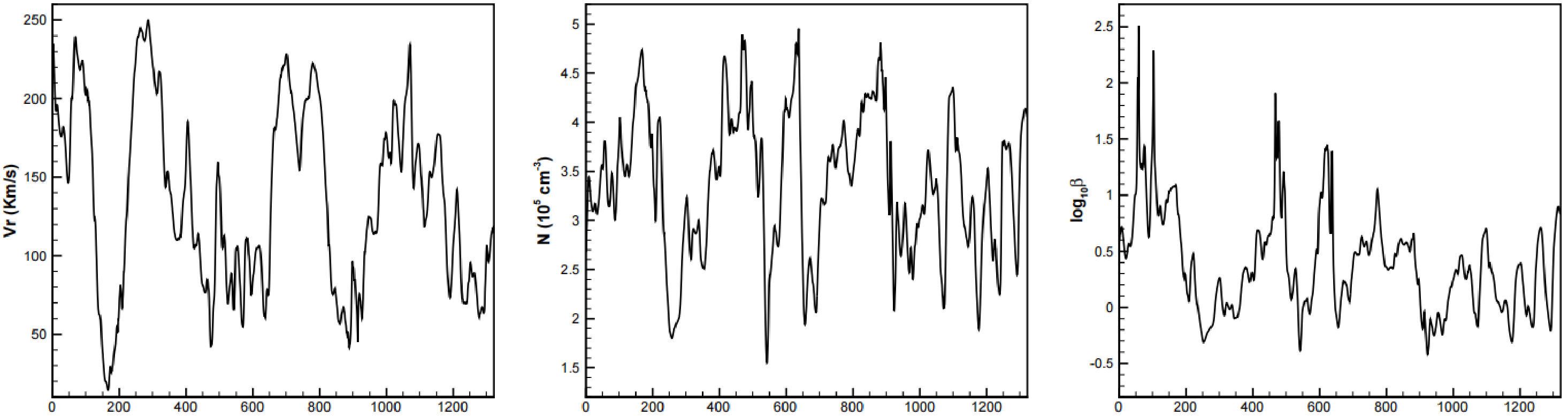}
\end{center}
\caption{Timing diagram of radial velocity $V_r$ ($\rm km~s^{-1}$, left), proton number density ($\rm 10^5 ~ cm^{-3}$, middle), and decadic logarithms of plasma $\beta$ (right), as observed by a virtual satellite located at 3 $R_s$, positioned $60^\circ$ behind Earth in longitude and at the same latitude.}\label{VrNBetaat3Rs}
\end{figure}

During the time-evolving coronal simulation, two virtual satellites are placed at 3 $R_s$ and 20 $R_s$, maintaining the same latitude as Earth and lagging $60^\circ$ behind in longitude. In Fig.~\ref{VrNBetaat3Rs}, we present the timing diagrams of radial velocity $V_r$ ($\rm km~s^{-1}$, left), proton number density ($\rm 10^5 ~ cm^{-3}$, middle), and decadic logarithms of plasma $\beta$ (right), as monitored by the virtual satellites positioned at 3 $R_s$. Fig.~\ref{VrNBetaat3Rs} demonstrate that the plasma $\beta$ at 3 $R_s$ around the solar equator usually varies between 0.3 to 100 during the time-evolving simulation, which is consistent with the values derived in \cite{Gary2001SoPh}. It is also observed that a peak$\big/$trough in plasma density profile generally corresponds to a trough$\big/$peak in the velocity profile. However, the peaks in plasma density around 600 and 675 hours are narrower than the corresponding troughs in the velocity profile, while the plasma density trough around 850 hours is wider than the corresponding velocity peak. These mismatches reveal the complexity of dynamic mechanisms in the subsonic$\big/$sub-Alfv{\'e}nic coronal region, highlighting the requirement of careful consideration of the transformation between kinetic and magnetic energy in simulations of such regions.

\subsection{Timing diagrams of plasma parameters at 20 $R_s$}\label{Insitu}
\begin{figure}[htpb]
\begin{center}
  \vspace*{0.01\textwidth}
    \includegraphics[width=0.8\linewidth,trim=1 1 1 1, clip]{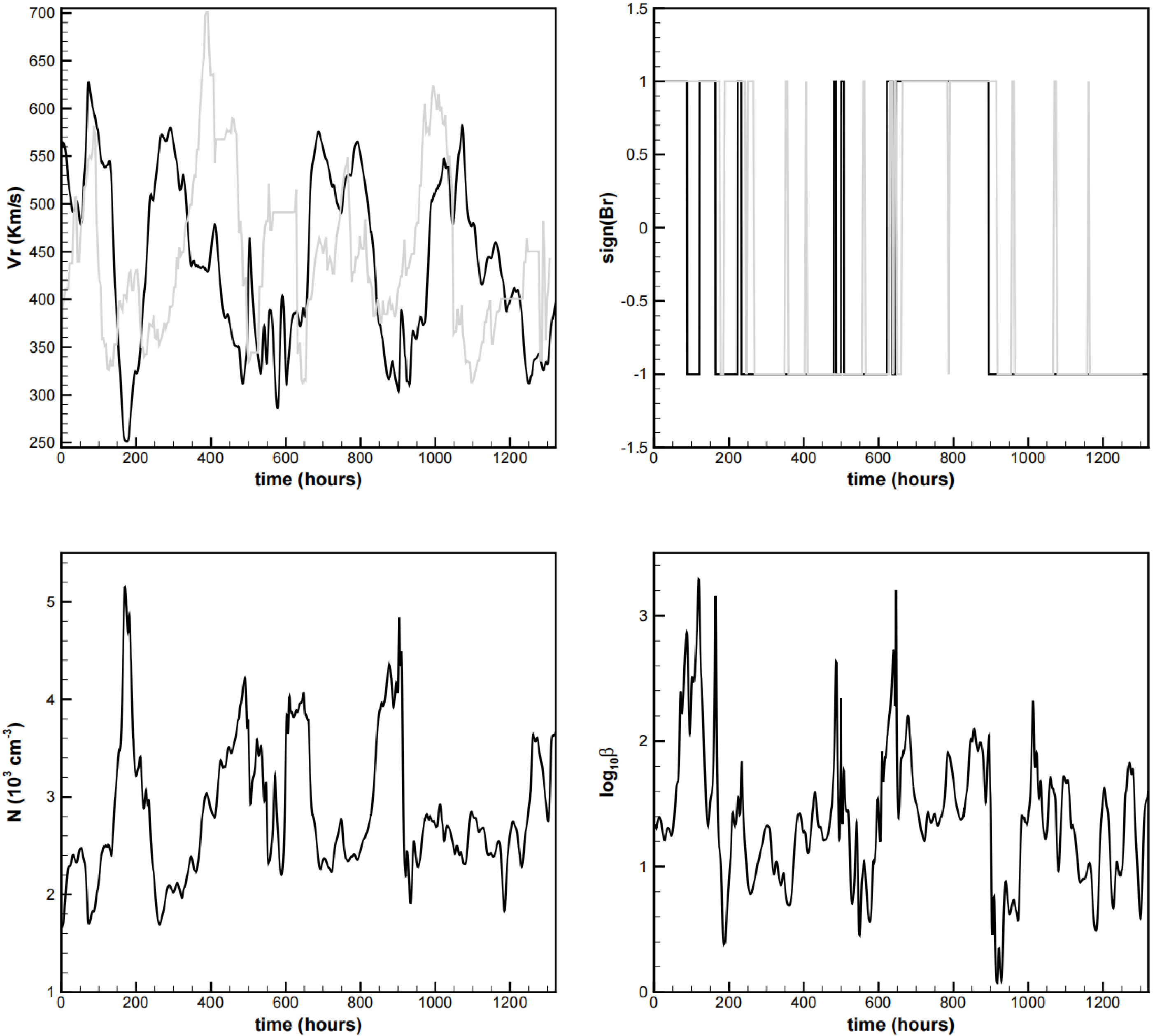}
\end{center}
\caption{Timing diagram of radial velocity $V_r$ ($\rm km~s^{-1}$, left top), proton number density ($\rm 10^3 ~ cm^{-3}$, left bottom), radial magnetic field polarities (right top) with ``-1" denoting the field pointing to the Sun and ``1" for the field directing away from the Sun, and decadic logarithms of plasma $\beta$ (right bottom). The black solid lines represent the simulated results observed by a virtual satellites located at 20 $R_s$, positioned $60^\circ$ behind Earth in longitude and at the same latitude. The gray solid lines denote the velocity and magnetic field polarity derived from WIND satellite observations, which has been mapped from 1 AU to 20 $R_s$ following the ballistic propagation.}\label{VrNBetaat20Rs}
\end{figure}

In Fig.~\ref{VrNBetaat20Rs}, we present the radial velocity $V_r$ ($\rm km~s^{-1}$, left top), proton number density ($\rm 10^3 ~ cm^{-3}$, left bottom), radial magnetic field polarities (right top), and decadic logarithms of plasma $\beta$ (right), as monitored by the virtual satellites positioned at 20 $R_s$ (black solid lines). Compared to Fig.~\ref{VrNBetaat3Rs}, the velocity peaks and troughs align more closely with plasma density troughs and peaks, indicating that the dynamics in this supersonic$\big/$super-Alfv{\'e}nic region are significantly simpler than those in the subsonic$\big/$sub-Alfv{\'e}nic coronal region.
Additionally, we noticed that the two velocity troughs appearing around 740.5 and 166.5 hours in Fig.~\ref{VrNBetaat3Rs} shifted to 748.2 and 175.0 hours in Fig.~\ref{VrNBetaat20Rs}, respectively. This indicates that the two troughs propagate from 3 $R_s$ to 20 $R_s$ with an average velocity of approximately 350 $\rm km~s^{-1}$. By performing a ballistic propagation, we map the solar wind velocity and radial magnetic field polarities observed by WIND satiate \citep{king2005JGR} at 1 AU to 20 $R_s$ (top, gray solid lines). Considering that some velocity peaks observed at 1 AU appear roughly 100 hours earlier when mapped to 20 $R_s$, we position the virtual satellite $60^\circ$ behind Earth to make a comparison with in-situ observations at 1 AU near Earth.

Fig.~\ref{VrNBetaat20Rs} shows that the simulation captures the observed velocity peak around 73 hours and 750 hours. The simulated velocity peak, approximately 580 $\rm km~s^{-1}$, occurs at 280 hours in the simulation, while it appears at 410 hours in the observations. Additionally, the velocity exceeding 580 $\rm km~s^{-1}$, which persists between 370 and 410 hours in the observations, is missing in the simulation. The observed velocity peak around 1000 hours appears about 70 hours earlier than in simulation. Additionally, the simulation approximately captures the observed velocity troughs around 180 hours, 880 hours. The two observed velocity troughs, centered at 500 and 650 hours, combine into one wide trough in the simulation, while the observed trough centered at 1100 hours occurs about 150 hours earlier than in the simulation. As for the radial magnetic polarities, the simulation capture $83.4 \%$ of the observations.
These discrepancies in velocity and radial magnetic polarities may be attributed to the limitations of the empirical heating function used to approximate coronal heating and solar wind acceleration, as well as the constraints of synoptic magnetograms, where the magnetic field at different longitudes is observed at different times. Since this period occurs during solar maximum, the frequent eruptions, such as coronal mass ejections, which are not included in this model, may also contribute to the discrepancies between the simulation results and observations.

\begin{figure}[htpb]
\begin{center}
  \vspace*{0.01\textwidth}
    \includegraphics[width=0.6\linewidth,trim=1 1 1 1, clip]{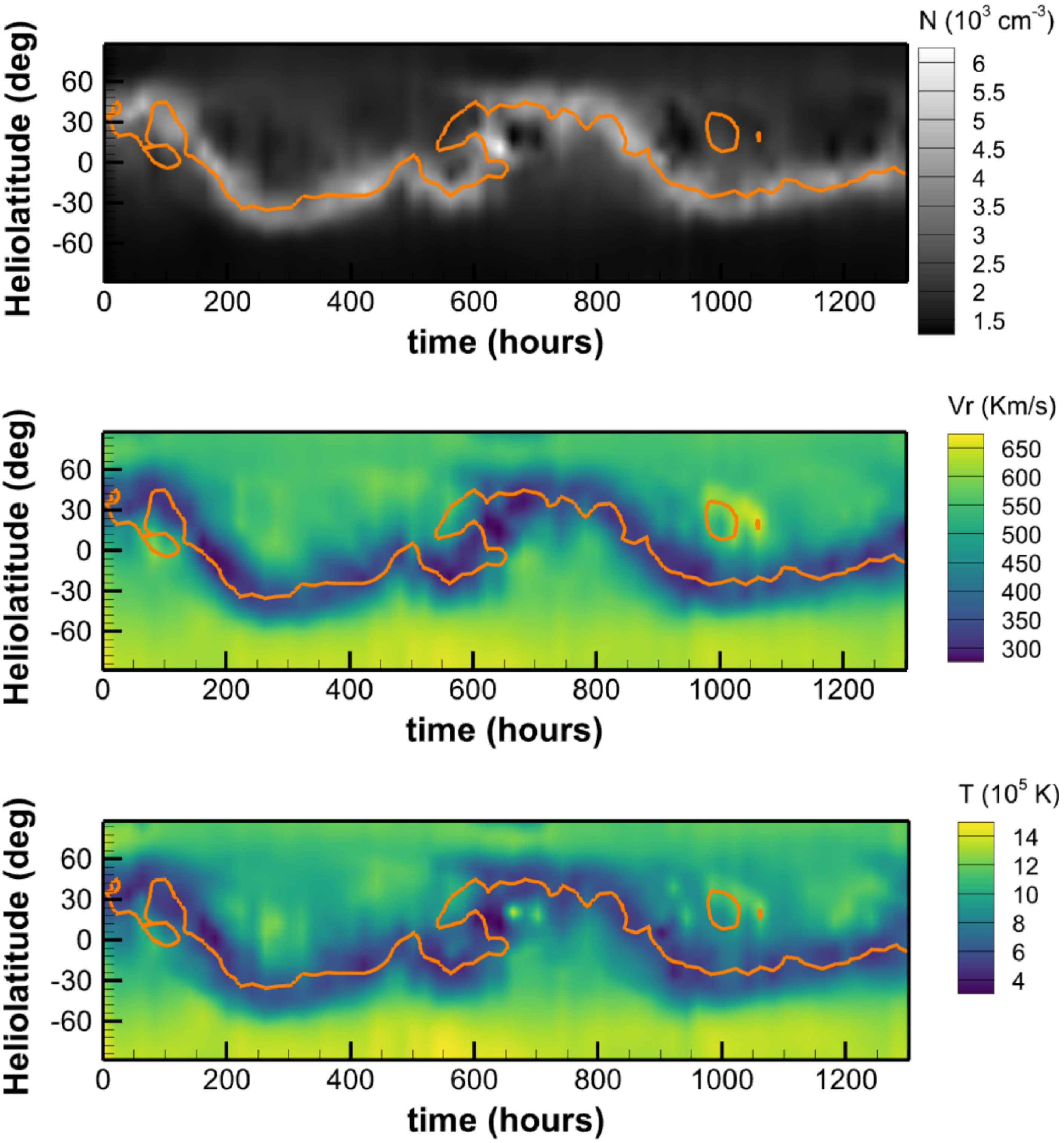}
\end{center}
\caption{Timing diagrams of simulated plasma number density ($\rm 10^3 ~ cm^{-3}$, top), radial velocity $V_r$ ($\rm km~s^{-1}$, middle) and temperature ($\rm 10^{5}~K$, bottom) at 20 $R_s$, corresponding to the same longitude as in Fig. \ref{VrNBetaat20Rs}.}\label{PannelNandVrat20Rs}
\end{figure}

In Fig.~\ref{PannelNandVrat20Rs}, we further present the 2 D timing diagrams of simulated plasma number density ($\rm 10^3 ~ cm^{-3}$, top), radial velocity $V_r$ ($\rm km~s^{-1}$, middle) and temperature ($\rm 10^{5}~K$, bottom) to show if the code produces the latitude structure of the solar wind, which is basically the latitude distribution of fast and solar wind. These diagrams are synthesized from a series of time-evolving simulation results at the same longitude as the virtual satellite, with a cadence of one result every 20 hours. The orange solid lines overlaid on the contours represent the MNLs. It shows that the MNLs, spanning between $30^{\circ} {\rm S}$ and $45^{\circ} {\rm S}$, are flatter than those in Fig.~\ref{pBat3Rs}. The low speed solar wind ($V_r$ < 400 $\rm km~s^{-1}$), associated with high plasma density, is primarily concentrated around the MNLs. Meanwhile, the fast solar wind ($V_r$ > 550 $\rm km~s^{-1}$), accompanied with low plasma density, dominate the polar regions south of $50^{\circ} {\rm S}$ and north of $80^{\circ} {\rm N}$. The simulated plasma temperature distributions are positively correlated with radial solar wind speeds \citep{Elliott2012,Pinto_2017,Licaixia2018}, with the fast solar wind exhibiting temperatures ranging from 10 to 14.7 M K, while the slow wind corresponds to temperatures between 3 and 7 M K. This demonstrates that the simulated temperature range match observations.

\section{Concluding remarks}\label{sec:Conclusion}
In this work, we extent our recently developed SIP-IFVM \citep{WangSubmitted} model, an implicit MHD coronal model constrained by a time-invariant magnetogram, into a time-evolving coronal model and adjust the adiabatic index $\gamma$ from 1.05 to $\frac{5}{3}$ to better represent the adiabatic process in coronal simulations. Furthermore, we design a magnetic field decomposition strategy for time-evolving coronal simulations and apply it to the time-evolving SIP-IFVM coronal model. Unlike traditional decomposition strategies, which split the magnetic field into a time-invariant potential field and a time-dependent component $\mathbf{B}_1$, failing to maintain $\mathbf{B}_1$ small during time-evolving simulations, our approach ensures that $\mathbf{B}_1$ remains consistently small throughout the simulations by introducing a temporally piecewise-constant variable to accommodate part of the non-potential field. As a result, the occurrence of non-physical negative thermal pressure when deriving thermal pressure from energy density in low-$\beta$ regions is effectively mitigated in time-evolving coronal simulations.

The main contribution of this paper is demonstrating that, by employing an implicit algorithm and adopting the novel extended magnetic field decomposition strategy proposed herein, we can perform 3 D global MHD coronal simulations to evaluate the long-term evolution of coronal structures in practical applications, achieving a speed more than 80 times faster than the real time evolution using only 192 CPU cores. By this means, the time-evolving MHD coronal model, driven by a series of time-evolving magnetograms, can robustly and efficiently resolve low-$\beta$ issues. Retaining more critical time-evolving information than commonly used quasi-steady-state coronal models further enables it to capture the dynamic features of the corona with higher fidelity and more accurately simulate solar disturbances, such as CME propagations.

We employ the thermodynamic time-evolving MHD coronal model, incorporating the extended magnetic field decomposition strategy, to simulate the evolution of the global corona from the solar surface to 20 $R_s$ during two solar maximum CRs with in an inertial coordinate system. Completing the 1300-hour evolution of the corona within just 16 hours of computational time validate its efficiency. Given that the maximum magnetic field strength near the solar surface occasionally exceeds 40 Gauss and the plasma beta reaches a minimum of approximately $10^{-3}$, the model demonstrates sufficient numerical stability for most solar-terrestrial simulation requirements. Furthermore, the time-evolving simulation results basically reproduce remote EUV and pB observations and capture in-situ measurements mapped from 1 AU to 20 $R_s$, demonstrating its capability to simulate complex time-evolving coronal structures during solar maximum.

Given that the computational time for simulating 1300 hours of physical time is no more than 16 hours using 192 CPU cores, it is practical to conduct faster-than-real-time CME simulations from the solar surface to 1 AU based on this work. We will refine the mesh and extend the coronal model to 1 AU, establishing a 3D implicit time-evolving MHD Sun-to-Earth model chain. Additionally, we will use an observation-based flux rope to trigger a realistic CME event in the time-evolving solar-terrestrial MHD model, validating this new generation of CME simulations. Furthermore, this model will be used to enhance and streamline the daily Sun-to-Earth forecasting process and to investigate the dynamic interaction between the solar wind and the magnetosphere of planets like Jupiter and Saturn.

Although this fully implicit time-evolving MHD solar coronal model, integrated with the extended magnetic field decomposition strategy, offers significant advantages and is a promising tool for timely and accurate simulation of the time-evolving corona in practical space weather forecasting, there is still considerable room for further improvement. 
Synchronized magnetograms are needed to address the limitation of current synoptic magnetographs, where magnetic fields at different longitudes are observed at different times. More accurate measurements of the photospheric magnetic fields in the polar regions are required to reproduce more realistic coronal structures. Additionally, more physically consistent heating source terms are needed to better mimic coronal heating and solar wind acceleration during time-evolving simulations. Moreover, self-consistently simulating the formation and evolution of coronal eruptions is crucial for enhancing the reliability of space weather forecasting using numerical models.

In our future work, we plan to incorporate surface flux transport models, such as the Advective Flux Transport (AFT) model \citep{Upton2014a}, to advect the radial magnetic field with observed flows and generate a more realistic magnetic field evolution at the inner boundary of our coronal model. We may also explore the use of artificial intelligence (AI) methods to enhance the resolution of the magnetic field in the polar regions, which are often poorly resolved in observations. Some local active region models \citep[e.g.][]{Amari2018,Jiang201605,Zhong2021} will be integrated into this global MHD coronal model to generate energized fields corresponding to transient eruption events. Additionally, we may incorporate additional observations into the inner boundary conditions, such as inferring horizontal velocities from observational data using the time-distance helioseismology method \citep{Yalim_2017,Zhao2012}. Furthermore, we plan to investigate the wave$\big/$turbulence-driven (WTD) heating mechanism \citep{Cranmer_2010,Schleich2023} in our time-evolving coronal model to gain a better understanding of how both fast and slow solar wind streams are heated and accelerated.

\vspace{0.5cm}
%\begin{acknowledgments}
\raggedright
$Acknowledgements.$ \quad This project has received funding from the European Research Council Executive Agency (ERCEA) under the ERC-AdG agreement No. 101141362 (Open SESAME).
These results were also obtained in the framework of the projects FA9550-18-1-0093 (AFOSR), C16/24/010  (C1 project Internal Funds KU Leuven), G0B5823N and G002523N (WEAVE) (FWO-Vlaanderen), and 4000145223 (SIDC Data Exploitation (SIDEX), ESA Prodex). This work also benefits from the National Natural Science Foundation of China (grant Nos. 42030204 and 42204155).
The resources and services used in this work were provided by the VSC (Flemish Supercomputer Centre), funded by the Research Foundation – Flanders (FWO) and the Flemish Government.
This work utilises data obtained by the Global Oscillation Network Group (GONG) program, managed by the National Solar Observatory and operated by AURA, Inc., under a cooperative agreement with the National Science Foundation. The data were acquired by instruments operated by the Big Bear Solar Observatory, High Altitude Observatory, Learmonth Solar Observatory, Udaipur Solar Observatory, Instituto de Astrof{\'i}sica de Canarias, and Cerro Tololo Inter-American Observatory. The authors also acknowledge the use of the STEREO/SECCHI data produced by a consortium of the NRL (US), LMSAL (US), NASA/GSFC (US), RAL (UK), UBHAM (UK), MPS (Germany), CSL (Belgium), IOTA (France), and IAS (France).
%\end{acknowledgments}

\appendix
\section{Derivation of the extended decomposed MHD equations}\label{DerivationofthedecomposedMHDequations}
Considering that $\left(\mathbf{B}+\epsilon~\mathbf{B}\right)^2-\mathbf{B}^2\equiv 2~\epsilon~\mathbf{B}^2+\epsilon^2~\mathbf{B}^2$ with $\epsilon~\mathbf{B}$ denoting the discretization error in the magnetic field $\mathbf{B}$, the magnetic pressure discretization error can be comparable to thermal pressure in low $\beta$ (the ratio of the thermal pressure to the magnetic pressure) regions and non-physical negative thermal pressure are prone to appear when deriving thermal pressure from energy density.
To avoid such undesirable situations, traditional decomposed MHD equations are commonly used, where the magnetic field $\mathbf{B}$ is divided into a time-independent potential field and a time-dependent field $\mathbf{B}_1$, with $\mathbf{B}_1$ being used in the derivation of the thermal pressure. However, discretization errors in $\mathbf{B}_1$ are still likely to result in non-physical negative thermal pressure as $\left|\mathbf{B}_1\right|$ increases in the time-evolving simulations, potentially causing the code to break down. Therefore, we propose the extended magnetic field decomposition strategy, in which the magnetic field $\mathbf{B}$ is divided into a time-independent potential field $\mathbf{B}_{00}$, a temporally piecewise constant field $\mathbf{B}_{01}$, and a time-dependent field $\mathbf{B}_1$.
In the following we describe how the extended decomposed MHD equations are derived.

With $\mathbf{B}_{00}$ representing  a static potential field, and $\mathbf{B}_{01}$ a temporally piecewise constant field that remains unchanged during the time interval between $t^n$ and $t^{n+k}$, where the subscripts ``$^n$" and ``$^{n+k}$" denote the $n$-th and $\left(n+k\right)$-th time levels in a simulation, the following conditions are satisfied within the time interval between $t^n$ and $t^{n+k}$:
\begin{equation}\label{PFB0}
\left\{\begin{array}{c}
\frac{\partial\mathbf{B}_{00}}{\partial t}=\mathbf{0}, ~ \nabla \cdot \mathbf{B}_{00}=0, ~ \nabla \times \mathbf{B}_{00}=0, ~ t \in [t^n, t^{n+k}] \\
\frac{\partial\mathbf{B}_{01}}{\partial t}=\mathbf{0}, ~ t \in (t^n, t^{n+k})
\end{array}\right..\
\end{equation}

The original energy equation is described as
\begin{equation}\label{orienergyequation}
\frac{\partial E}{\partial t}+\nabla \cdot\left[\left(E+p_{T}\right)\mathbf{v}-\mathbf{B}\left(\mathbf{v}\cdot\mathbf{B}\right)\right]
=-\left(\nabla \cdot \mathbf{B}\right)\mathbf{v} \cdot \mathbf{B}
\end{equation}
where $E=\frac{p}{\gamma-1}+\frac{1}{2}\rho\mathbf{v}^{2}+\frac{1}{2}\mathbf{B}^{2}$ and $p_{T}=p+\frac{\mathbf{B}^2}{2}$. Given that $\mathbf{B}=\mathbf{B}_{00}+\mathbf{B}_{01}+\mathbf{B}_1$ and $E_1=\frac{p}{\gamma-1}+\frac{1}{2}\rho \, \mathbf{v}^{2}+\frac{1}{2}\mathbf{B}_1^{2}$.
The formulation of $E$ can be described as
\begin{equation}\label{OriEnergy}
E=E_1+\frac{1}{2}\mathbf{B}_{0}^2+\mathbf{B}_0\cdot\mathbf{B}_1
\end{equation}
where $\mathbf{B}_0=\mathbf{B}_{00}+\mathbf{B}_{01}$.

The original induction equation is described as
\begin{equation}\label{oriinductionequation}
\frac{\partial \mathbf{B}}{\partial t}+\nabla \cdot\left(\mathbf{v \, B}-\mathbf{B \, v}\right)=-\left(\nabla \cdot \mathbf{B}\right)\mathbf{v}
\end{equation}

From Eq.~(\ref{PFB0}) and Eq.~(\ref{OriEnergy}) we get
\begin{equation}\label{EnergyEquation}
\frac{\partial E}{\partial t}=\frac{\partial E_1}{\partial t}+ \mathbf{B}_{0}\cdot \frac{\partial \mathbf{B}_0}{\partial t} + \mathbf{B}_0 \cdot \frac{\partial \mathbf{B}_1}{\partial t}   + \mathbf{B}_1 \cdot \frac{\partial \mathbf{B}_0}{\partial t}
\end{equation}
\begin{equation}\label{PartialB}
\frac{\partial \mathbf{B}}{\partial t}=\frac{\partial  \left(\mathbf{B}_1+\mathbf{B}_{01}\right)}{\partial t}
\end{equation}

From Eq.~(\ref{PartialB}) and Eq.~(\ref{oriinductionequation}) we get
\begin{equation}\label{PartialB1}
\frac{\partial \mathbf{B}_1}{\partial t} = -\nabla \cdot \left( \mathbf{v} \, \mathbf{B} - \mathbf{B} \, \mathbf{v} \right) - \mathbf{v} \, \nabla \cdot \mathbf{B} -\frac{\partial \mathbf{B}_{01}}{\partial t}
\end{equation}
Multiplying both the left-hand side and the right-hand side of Eq.~(\ref{PartialB1}) by $\mathbf{B}_0$ results in
\begin{equation}\label{B0dotPartialB1_ori}
\mathbf{B}_0 \cdot \frac{\partial \mathbf{B}_1}{\partial t} = -\nabla \cdot \left( \mathbf{v} \, \mathbf{B} - \mathbf{B} \, \mathbf{v} \right) \cdot \mathbf{B}_0 - \mathbf{v} \cdot \mathbf{B}_0 \, \nabla \cdot \mathbf{B} -\mathbf{B}_0 \cdot \frac{\partial \mathbf{B}_{01}}{\partial t}
\end{equation}
Considering that
$$\nabla \cdot \left( \mathbf{v} \, \mathbf{B}\right)\cdot \mathbf{B}_0=\left( \nabla \cdot \mathbf{v}\right) \mathbf{B} \cdot \mathbf{B}_0+ \left(\mathbf{v} \cdot \nabla \right)\mathbf{B} \cdot \mathbf{B}_0 = \nabla \cdot \left[ \mathbf{v} \, \left(\mathbf{B} \cdot \mathbf{B}_0\right)\right]
-\left(\mathbf{v} \cdot \nabla \right)\mathbf{B}_0 \cdot \mathbf{B}$$
$$\nabla \cdot \left( \mathbf{B} \, \mathbf{v}\right)\cdot \mathbf{B}_0=\left( \nabla \cdot \mathbf{B}\right) \mathbf{v} \cdot \mathbf{B}_0+ \left(\mathbf{B} \cdot \nabla \right)\mathbf{v} \cdot \mathbf{B}_0 = \nabla \cdot \left[ \mathbf{B} \, \left(\mathbf{v} \cdot \mathbf{B}_0\right)\right]
-\left(\mathbf{B} \cdot \nabla \right)\mathbf{B}_0 \cdot \mathbf{v}$$
Eq.~(\ref{B0dotPartialB1_ori}) is equivalent to
\begin{equation}\label{B0dotPartialB1}
\mathbf{B}_0 \cdot \frac{\partial \mathbf{B}_1}{\partial t} = -\nabla \cdot \left[ \left(\mathbf{B} \cdot \mathbf{B}_0\right) \, \mathbf{v}  - \left(\mathbf{v} \cdot \mathbf{B}_0\right) \, \mathbf{B} \right] - \mathbf{v} \cdot \mathbf{B}_0 \, \nabla \cdot \mathbf{B} -\mathbf{B}_0 \cdot \frac{\partial \mathbf{B}_{01}}{\partial t}
+\eta
\end{equation}
where $\eta=\left(\mathbf{v} \cdot \nabla \right)\mathbf{B}_0 \cdot \mathbf{B}-\left(\mathbf{B} \cdot \nabla \right)\mathbf{B}_0 \cdot \mathbf{v}$.

From Eq.~(\ref{B0dotPartialB1}) and Eq.~(\ref{EnergyEquation}) we get
\begin{equation}\label{partialE}
\frac{\partial E}{\partial t} = \frac{\partial E_1}{\partial t} - \nabla \cdot \left[ \left(\mathbf{B} \cdot \mathbf{B}_0\right) \, \mathbf{v}  - \left(\mathbf{v} \cdot \mathbf{B}_0 \right) \, \mathbf{B} \right] - \mathbf{v} \cdot \mathbf{B}_0 \, \nabla \cdot \mathbf{B} -\mathbf{B}_0 \cdot \frac{\partial \mathbf{B}_{01}}{\partial t}  +
\mathbf{B}_{0}\cdot \frac{\partial \mathbf{B}_0}{\partial t}   + \mathbf{B}_1 \cdot \frac{\partial \mathbf{B}_0}{\partial t}
+\eta
\end{equation}

From Eq.~(\ref{orienergyequation}) and Eq.~(\ref{OriEnergy}) we get
\begin{equation}\label{partialEandB1}
\frac{\partial E}{\partial t} = -\nabla \cdot \left[ \left( E_1 + p + \mathbf{B}_0^2 + 2 \, \mathbf{B}_0 \cdot \mathbf{B}_1 + \frac{1}{2} \, \mathbf{B}_1^2 \right) \mathbf{v} - (\mathbf{v} \cdot \mathbf{B}) \mathbf{B} \right] - \mathbf{v} \cdot \mathbf{B}  \, \nabla \cdot \mathbf{B}
\end{equation}

From Eq.~(\ref{partialEandB1}) and Eq.~(\ref{partialE}) we get
\begin{equation}\label{partialE1andB1}
\begin{aligned}
\frac{\partial E_1}{\partial t} &+ \nabla \cdot \left[ \left( E_1 + p + \mathbf{B}_0 \cdot \mathbf{B}_1 + \frac{1}{2} \, \mathbf{B}_1^2 \right) \mathbf{v} - (\mathbf{v} \cdot \mathbf{B}_1) \mathbf{B} \right] \\& = -\mathbf{v} \cdot \mathbf{B}_1 \, \nabla \cdot \mathbf{B} +\mathbf{B}_0 \cdot \frac{\partial \mathbf{B}_{01}}{\partial t}
-\left(\mathbf{B}_{0}\cdot \frac{\partial \mathbf{B}_0}{\partial t}   + \mathbf{B}_1 \cdot \frac{\partial \mathbf{B}_0}{\partial t} \right)
-\eta
\end{aligned}
\end{equation}

Since $\nabla \cdot \mathbf{B}_{00} = 0$, we obtain the following decomposed energy equation:
\begin{equation}\label{decomposedenergyequation}
\begin{aligned}
\frac{\partial E_1}{\partial t} &+\nabla \cdot\left[\left(E_1+p_{T1}+\mathbf{B}_1\cdot\mathbf{B}_0\right)\mathbf{v}-\mathbf{B}\left(\mathbf{v}\cdot\mathbf{B}_1\right)\right]
=\\& -\mathbf{v} \cdot \mathbf{B}_1 \nabla \cdot \left(\mathbf{B}_1+\mathbf{B}_{01} \right) +\mathbf{B}_0 \cdot \frac{\partial \mathbf{B}_{01}}{\partial t}
-\left(\mathbf{B}_{0}\cdot \frac{\partial \mathbf{B}_0}{\partial t}   + \mathbf{B}_1 \cdot \frac{\partial \mathbf{B}_0}{\partial t} \right)
-\eta
\end{aligned}
\end{equation}
where $p_{T1}=p+\frac{\mathbf{B}_1^2}{2}$.
This indicates that the thermal pressure, $p$, can be determined from $E_1$. As a result, the accuracy of $p$ is no longer limited by the discretization error of $\mathbf{B}$ but rather by that of $\mathbf{B}_1$. As long as $\mathbf{B}_1$ remains small, the discretization error in $\frac{1}{2} \, \mathbf{B}_1^2$ is unlikely to approach the magnitude of the thermal pressure, reducing the risk of non-physical negative thermal pressure.

The original momentum equation is described as
\begin{equation}\label{MomentumEq}
\frac{\partial\left(\rho \mathbf{v}\right)}{\partial t}+\nabla \cdot\left[\rho\mathbf{v} \, \mathbf{v}+\left(p+\frac{\mathbf{B}^2}{2}\right)\mathbf{I}- \mathbf{B} \, \mathbf{B}\right] = - \left(\nabla \cdot \mathbf{B}\right) \mathbf{B}
\end{equation}

Given that $\nabla \times \mathbf{B}_{00} \times \mathbf{B}_{00}=\nabla \cdot \left(-\frac{1}{2}\mathbf{B}_{00}^2\mathbf{I}+\mathbf{B}_{00} \, \mathbf{B}_{00}\right)-\left(\nabla \cdot \mathbf{B}_{00}\right) \mathbf{B}_{00}$ and $\nabla \cdot \mathbf{B}_{00} = 0$, Eq.~(\ref{MomentumEq}) is equivalent to
\begin{equation}\label{decomposedMomentumEq}
\frac{\partial (\rho \, \mathbf{v})}{\partial t} + \nabla \cdot \left[ \rho \mathbf{v \, v} + \left( p + \frac{\mathbf{B}^2}{2} - \frac{\mathbf{B}_{00}^2}{2} \right) \mathbf{I} - \mathbf{B \, B} + \mathbf{B}_{00} \, \mathbf{B}_{00} \right] = - \nabla \cdot \left( \mathbf{B}_1 + \mathbf{B}_{01} \right) \mathbf{B}
\end{equation}

Consequently, we get the following extended decomposed MHD equations:
\begin{equation}\label{ExtendedDecomposision}
\left\{
\begin{array}{l}
\frac{\partial \rho}{\partial t} + \nabla \cdot (\rho \, \mathbf{v}) = 0 \\
\frac{\partial (\rho \, \mathbf{v})}{\partial t} + \nabla \cdot \left[ \rho \mathbf{v \, v} + \left( p + \frac{\mathbf{B}^2}{2} - \frac{\mathbf{B}_{00}^2}{2} \right) \mathbf{I} - \mathbf{B \, B} + \mathbf{B}_{00} \, \mathbf{B}_{00} \right] = - \nabla \cdot \left( \mathbf{B}_1 + \mathbf{B}_{01} \right) \mathbf{B} \\
\frac{\partial E_1}{\partial t} + \nabla \cdot \left[ \left( E_1 + p_{T1} + \mathbf{B}_1 \cdot \mathbf{B}_0 \right) \mathbf{v} - \mathbf{B} \left( \mathbf{v} \cdot \mathbf{B}_1 \right) \right] =  - \nabla \cdot \left( \mathbf{B}_1 + \mathbf{B}_{01} \right) \left( \mathbf{v} \cdot \mathbf{B}_1 \right) \\
\quad +\mathbf{B}_0 \cdot \frac{\partial \mathbf{B}_{01}}{\partial t}  -\left(\mathbf{B}_{0}\cdot \frac{\partial \mathbf{B}_0}{\partial t}   + \mathbf{B}_1 \cdot \frac{\partial \mathbf{B}_0}{\partial t} \right)
-\left(\mathbf{v} \cdot \nabla \right)\mathbf{B}_0 \cdot \mathbf{B}+\left(\mathbf{B} \cdot \nabla \right)\mathbf{B}_0 \cdot \mathbf{v}\\
\frac{\partial \mathbf{B}_1}{\partial t} + \nabla \cdot (\mathbf{v \, B} - \mathbf{B \, v}) = - \nabla \cdot \left( \mathbf{B}_1 + \mathbf{B}_{01} \right) \mathbf{v}-\frac{\partial \mathbf{B}_{01}}{\partial t}
\end{array}
\right.
\end{equation}

Suppose $\mathbf{B}_{01}$ is increased by $\mathbf{B}_1$ over a very short period when $\mathbf{B}_1$ becomes significantly large at time $t^{n}$, such as when $\frac{p}{0.5 \, \mathbf{B}_1^2}$ drops below a specific threshold. In this case, the terms $\nabla \cdot \left(\mathbf{v \, B} - \mathbf{B \, v}\right)$ and $- \nabla \cdot \left( \mathbf{B}_1\right)$ become negligible in Eq. (\ref{PartialB1}) and $\frac{\partial \mathbf{B}_1}{\partial t}=-\frac{\partial \mathbf{B}_{01}}{\partial t}$ holds. Consequently, $\mathbf{B}_1$ decreases to $\mathbf{0}$ over this very short period, $\mathbf{B}_0 \cdot \frac{\partial \mathbf{B}_{01}}{\partial t}  -\left(\mathbf{B}_{0}\cdot \frac{\partial \mathbf{B}_0}{\partial t}   + \mathbf{B}_1 \cdot \frac{\partial \mathbf{B}_0}{\partial t} \right)$ is equivalent to $\mathbf{B}_1 \cdot \frac{\partial \mathbf{B}_{1}}{\partial t}$, and Eq. (\ref{decomposedenergyequation}) reduces to $\frac{\partial E_1}{\partial t}= \frac{\partial \left(\frac{1}{2} \mathbf{B}_1^2\right)}{\partial t}$. At the end of this infinitely short period, $\mathbf{B}_{01}$, $\mathbf{B}_1$ and $E_1$ become $\mathbf{B}_{01}+\mathbf{B}_1$, $\mathbf{0}$, and $\frac{p}{\gamma-1}+\frac{1}{2}\rho\mathbf{v}^{2}$, respectively.

Therefore, when $\frac{p}{0.5 \, \mathbf{B}_1^2}$ drops below a threshold at time $t^{n}$, we update $\mathbf{B}_{01}$ to $\mathbf{B}_{01}+\mathbf{B}_1$, set $\mathbf{B}_1=0$, and assign $E_1=\frac{p}{\gamma-1}+\frac{1}{2}\rho\mathbf{v}^{2}$. Subsequently, we solve Eq.~(\ref{ExtendedDecomposision}) with $\frac{\partial \mathbf{B}_{01}}{\partial t}=0$ to advance the solutions in time. This process continues until $\frac{p}{0.5 \, \mathbf{B}_1^2}$ once again drops below the threshold, at which moment we repeat the procedure outlined above.

\bibliographystyle{spr-mp-sola}
\bibliography{SIPIFVMB00B01}



\end{document}